\documentclass[preprint,12pt]{elsarticle}
\usepackage{algorithm}
\usepackage{algorithmic}
\usepackage{float}
\usepackage{xcolor}
\usepackage{amssymb}
\usepackage{bm}
\usepackage{amsmath}
\usepackage{dsfont}
\usepackage{booktabs}
\usepackage{multirow}
\usepackage{graphicx}
\usepackage{hyperref}
\usepackage{cleveref}
\usepackage{enumitem}

\usepackage{amssymb}
\usepackage{amsmath}

\journal{JMLR}

\begin{document}

\begin{frontmatter}



\title{EEGDancer: Dynamic Emotion Latent Space Masked Modeling with Reinforcement Learning for EEG Continuous Emotion Prediction}
\tnotetext[equal]{\textsuperscript{1}Equal contributions.}


\author{Zhihao Zhou\textsuperscript{a,b},
Weishan Ye\textsuperscript{a,b}, Li Zhang\textsuperscript{a,b}, Gan Huang\textsuperscript{a,b}, Zhen Liang\textsuperscript{a,b,*}} 

\affiliation[a]{organization={The School of Biomedical Engineering},
            addressline={Medical School}, 
            city={Shenzhen University},
            state={Shenzhen},
            country={China}}
\affiliation[b]{organization={The Guangdong Provincial Key Laboratory of Biomedical Measurements and Ultrasound Imaging},
            state={Shenzhen},
            country={China}}

\begin{abstract}
{Continuous electroencephalography (EEG) emotion prediction aims to model the dynamic temporal evolution of human emotional states from EEG signals. Unlike conventional discrete emotion recognition, continuous emotion prediction requires the model to capture long-range temporal dependencies and globally coherent emotional evolution patterns. However, existing methods mainly rely on point-wise supervised regression and directly model noisy high-dimensional EEG features, making it difficult to effectively characterize continuous emotional dynamics. To address these challenges, this paper proposes EEGDancer, a dynamic emotional latent space learning framework for continuous EEG emotion prediction. Similar to how humans naturally synchronize with musical rhythms, EEGDancer learns to align with the temporal dynamics of EEG emotional signals and continuously adapt to emotional state variations over time. The proposed framework jointly integrates vector-quantized representation learning, masked temporal modeling, and reinforcement learning-based trajectory optimization into a unified architecture.Specifically, a causal spatiotemporal Vector-Quantization Variational Autoencoder (VQ-VAE) is introduced to learn structured emotional prototypes and construct a discrete-continuous emotional latent space from EEG signals. Based on the learned latent space, a Transformer-based masked dynamic modeling strategy is further employed to capture long-range emotional dependencies and intrinsic temporal evolution patterns. Furthermore, continuous emotion prediction is formulated as a sequential decision-making problem, and a Soft Actor–Critic (SAC) reinforcement learning framework is introduced to optimize emotional prediction trajectories at the sequence level rather than conventional frame-wise local fitting.Extensive experiments on the SEED, SEED-IV, and Long-Term Naturalistic Emotion datasets demonstrate that the proposed method consistently outperforms existing machine learning and deep learning approaches. Experimental and ablation results further verify the effectiveness of the proposed dynamic emotional latent space and reinforcement learning-based trajectory optimization for modeling continuous EEG emotional dynamics.The project page is available at: https://github.com/ZhaoZ77/EEGDancer.
}
\end{abstract}



\end{frontmatter}




\section{Introduction}
\label{sec:Introduction}

{Emotion plays a fundamental role in human cognition, decision-making, social interaction, and mental health. Accurate emotion recognition has therefore become an important research topic in affective computing, brain computer interfaces (BCIs)\cite{li2022eeg,alarcao2017emotions,houssein2022human}, healthcare, and human computer interaction. Among various physiological modalities, electroencephalography (EEG) has attracted extensive attention due to its non-invasive nature, high temporal resolution, and capability to directly reflect neural activities associated with emotional processing. Compared with external behavioral signals such as facial expressions or speech, EEG provides a more objective and reliable representation of intrinsic emotional states\cite{peng2022ogssl,pancholi2022source}.

With the rapid development of machine learning and deep learning techniques, EEG-based emotion recognition has achieved significant progress in recent years. Traditional approaches mainly relied on handcrafted temporal, spectral, and spatial features combined with conventional classifiers. More recently, deep neural networks, including convolutional neural networks\cite{assemlali2025deep} (CNNs), recurrent neural networks\cite{atlas2025modernized} (RNNs), graph neural networks\cite{yu2026ia} (GNNs), and Transformer-based\cite{wang2025cross} architectures, have demonstrated strong capability in automatically learning discriminative representations from EEG signals.

Despite these advances, most existing EEG emotion recognition studies formulate emotion analysis as a static classification problem, where an entire EEG trial is assigned a single discrete emotional label such as positive, neutral, or negative\cite{shu2018review}. However, human emotions are inherently dynamic and continuously evolving over time. During emotional experiences, emotional states exhibit gradual transitions, temporal continuity, and local fluctuations. Assigning a single label to a long EEG sequence inevitably ignores the intrinsic temporal evolution of emotional states and weakens the model’s ability to capture fine-grained emotional dynamics\cite{ji2022spatial}.

To address this issue, several recent studies have explored continuous affective regression and temporal sequence modeling. Transformer-based architectures have shown promising capability in capturing long-range temporal dependencies, while temporal regression frameworks have achieved encouraging performance in physiological signal analysis. Nevertheless, continuous EEG emotion prediction remains highly challenging due to three major limitations.}

{
First, EEG signals are highly non-stationary and noisy\cite{luo2026multi,liu2026cria}, making it difficult to learn stable emotional representations over long temporal horizons. Existing methods often directly model high-dimensional continuous EEG features, which are sensitive to irrelevant neural fluctuations and subject-specific variability.

Second, current methods usually lack an effective latent representation mechanism capable of jointly modeling discrete emotional semantics and continuous emotional transitions\cite{zhou2024emotvr}. Human emotional evolution typically follows smooth transitions between multiple latent emotional states rather than abrupt changes between isolated categories. However, conventional continuous representations often fail to preserve interpretable emotional structures, while purely discrete representations ignore gradual emotional evolution.

Furthermore, unlike conventional supervised regression, continuous EEG emotion prediction is inherently a sequential decision-making problem rather than an independent frame-wise prediction task\cite{ju2024eeg,zhou2025emotion}. Human emotional states exhibit strong temporal continuity and long-range dependencies, where the prediction at the current time step influences the consistency and evolution of future emotional trajectories. However, traditional supervised learning methods optimize only point-wise prediction errors and mainly focus on local fitting objectives, making them insufficient for modeling the global temporal evolution of emotional dynamics.

Moreover, EEG emotional signals are highly noisy and non-stationary. Optimizing solely with supervised regression losses often causes models to overfit instantaneous fluctuations while ignoring long-term emotional transition patterns. As a result, the predicted emotional trajectories may become temporally inconsistent and fail to reflect realistic emotional evolution processes.

To overcome these limitations, this paper proposes EEGDancer, a novel dynamic emotional latent space learning framework for continuous EEG emotion prediction. The proposed framework integrates vector-quantized representation learning, masked temporal modeling, and reinforcement learning into a unified architecture for dynamic EEG emotion modeling.Specifically, we first introduce a causal spatiotemporal Vector-Quantization Variational Autoencoder\cite{zhang2025mind,ma2025codebrain} (VQ-VAE) to learn a structured emotional codebook from unlabeled EEG signals in a self-supervised manner. The learned codebook captures representative emotional prototypes and provides discrete semantic priors for emotional dynamics modeling.Based on the pretrained emotional codebook, we further propose a Transformer-based\cite{bao2021beit,ren2025videoworld} masked dynamic modeling strategy to learn long-range temporal dependencies in a dynamic emotional latent space. Unlike conventional hard discrete representations, the proposed latent space represents emotional states as weighted combinations of multiple emotional prototypes, thereby simultaneously preserving semantic discreteness and temporal continuity.Finally, we formulate continuous emotion prediction as a sequential decision-making problem and further optimize temporal emotional prediction through a Soft Actor–Critic (SAC)\cite{haarnoja2018soft}  reinforcement learning framework. By introducing regression consistency and temporal continuity rewards, the proposed model effectively captures both local emotional transitions and global emotional evolution trends.Extensive experiments on the SEED, SEED-IV, and Long-Term Naturalistic Emotion datasets demonstrate that the proposed method consistently outperforms existing machine learning and deep learning approaches in continuous emotion prediction tasks.

The main contributions of this work are summarized as follows:

\begin{itemize}
    \item We propose EEGDancer, a unified dynamic emotional latent space learning framework for continuous EEG emotion prediction, which jointly integrates Vector-Quantization representation learning, masked temporal modeling, and reinforcement learning-based trajectory optimization.
    \item We introduce a causal spatiotemporal VQ-VAE together with a Transformer-based masked dynamic modeling strategy to construct a discrete-continuous emotional latent space, enabling the model to simultaneously capture structured emotional semantics, long-range temporal dependencies, and continuous emotional evolution patterns from EEG signals.
    \item We formulate continuous EEG emotion prediction as a sequential decision-making problem and further introduce a Soft Actor–Critic (SAC) reinforcement learning framework with temporal-aware reward optimization, allowing the model to perform trajectory-level optimization of emotional dynamics rather than conventional point-wise regression fitting.
    \item Extensive experiments and ablation studies conducted on the SEED, SEED-IV, and Long-Term Naturalistic Emotion datasets demonstrate that the proposed framework achieves superior performance, strong robustness, and effective temporal emotion tracking capability compared with existing machine learning and deep learning approaches.
    
\end{itemize}

\section{Related Work}
\subsection{{Reinforcement Learning}}
{Reinforcement Learning (RL) is a powerful paradigm in which an agent learns to make optimal decisions by interacting with an environment and receiving rewards \cite{zoph2017neuralarchitecturesearchreinforcement,guo2025deepseek}. Unlike static methods, RL continuously adapts its actions to maximize cumulative rewards, making them well-suited for dynamic and complex environments\cite{he2016deepreinforcementlearningnatural,yarats2021image}.

\subsection{Deep Reinforcement Learning in EEG}{
Deep Reinforcement Learning (DRL), offers significant promise for automating and enhancing EEG-based BCI tasks like emotion recognition, sleep staging and action imagery. 

To address the dynamic nature of EEG-based emotion recognition, Zhang et al. \cite{zhang2023unsupervised} proposed TAS-Net, an unsupervised time-aware sampling network via DRL. In contrast to traditional methods that rely on static, trial-level labels and ignore temporal dynamics, TAS-Net explicitly captured both short-term emotional continuity and long-range temporal dependencies. By identifying key emotional segments from continuous EEG signals, it enabled more informative feature extraction for downstream emotion recognition tasks. Experimental results showed that TAS-Net consistently outperforms existing unsupervised methods, offering a robust and effective solution for emotion-based BCI applications in dynamic and unlabeled environments.

To address the challenge of optimizing model structures, Yang et al. \cite{yang2021cnn} introduced a hierarchical DQN-PSO framework. In this approach, a Deep Q-Network (DQN) selected the core CNN architecture, while Particle Swarm Optimization (PSO) fine-tuned convolutional kernel parameters. This hybrid strategy enabled automated generation of high-performing EEG classification models. To address the limitations of CNNs in capturing the non-Euclidean spatial structure of EEG electrodes, Aung et al. \cite{aung2025real} proposed EEG RL-Net. This framework combined pretrained Graph Neural Networks (GNNs) for spatial feature extraction with a Dueling DQN module that dynamically skipped ambiguous temporal segments. The model improved motor imagery performance and supporting real-time applications. For applications involving limited clinical data, Liu et al.\cite{liu2023automatic} developed AFM-DQN, a framework that incorporated Semi-Joint Mutual Information (Semi-JMI) for feature selection within a DQN-based pipeline. The model also integrated a pretrained module, SVM classifier, and reward control mechanism to facilitate effective localization of epileptogenic zones. 

To overcome the constraints of explicit user feedback in reinforcement learning, EEG-derived Error-Related Potentials (ErrPs) is incoportated as implicit, low-latency reinforcement signals. For example, Xu et al.\cite{xu2021accelerating} demonstrated how ErrP-based feedback can reduce cognitive load and improve sample efficiency in real-time applications such as autonomous driving. Similarly, Luo et al. \cite{luo2018deep} embedded a CNN-decoded ErrP preference signal within a reinforcement teacher framework to enhance training in robotic systems. 

Inspired by neuroscience, Li et al. \cite{li2023brain} proposed FLD3QN, a biologically inspired DRL architecture that models emotional processing based on the Papez circuit and dopamine-driven reward mechanisms. This neurobiological grounding enabled significant improvements in emotion recognition accuracy, highlighting the potential of neuroscience-informed DRL designs in affective computing.
}
}

\subsection{{Vector Quantization}}
{The core of sparse representation learning lies in leveraging overcomplete dictionaries and sparse vectors to compress signals into compact combinations of basis atoms. Its theoretical foundation can be traced back to Shannon’s rate–distortion theory, which explains that under a given source distribution and distortion metric mean squared error (MSE), minimizing representation complexity with sparsity constraints enables approximation of the optimal distortion within a limited bit rate.In the field of signal processing\cite{zhang2025mind}, this efficiency-through-sparsity paradigm has demonstrated remarkable success. For instance, speech denoising benefits from precise capture of speech structures, significantly enhancing noise robustness; compressed sensing substantially improves signal reconstruction accuracy; Similarly, in image processing\cite{elad2006image}, sparse representation plays a crucial role in tasks such as denoising and super-resolution reconstruction\cite{yang2010image}.

Complementing the continuous optimization framework of sparse representation is vector quantization, a direct practical realization of rate–distortion theory. Vector quantization maps signals to the nearest codeword in a discrete codebook, with its essence lying in balancing distortion and efficiency under bit rate constraints\cite{gray1984vector}. In speech coding, vector quantization has redefined transmission standards through high compression efficiency\cite{makhoul1985vector}. In audio compression, discrete audio representation, and communication signal transmission, iterative codebook optimization further pushes the rate–distortion boundary toward its theoretical limits.

Therefore, sparse representation (continuous representation) and vector quantization (discrete representation), both guided by rate–distortion theory, jointly provide a solid theoretical foundation for tasks such as signal denoising and feature extraction\cite{mao2021discrete,aczel2026neural,zhao2026spherical}.}


\section{Methodology}{
\subsection{{Markov Decision Process Formulation}}
\label{subsec:Problem Definition}

{To detect the key emotional segments from sequential EEG signals, we formulate the task as a sequential decision-making problem modeled using a Markov Decision Process (MDP). An MDP is defined as a tuple $\mathcal{M}=\mathcal{<}\mathcal{S}, \mathcal{A}, \mathcal{P}, \mathcal{R}, \mathcal{\gamma} \mathcal{>}$. Here, the state space $\mathcal{S}$ represents various segments of the EEG signals, and the action space $\mathcal{A}=[-1,1]$ consists of two possible actions (action $1$ indicates the selection of a segment as emotionally relevant, and action $0$ denotes its rejection). The transition probability function $\mathcal{S} \times \mathcal{A} \times \mathcal{S} \rightarrow [0,1]$ describes the likelihood of transitioning from one state to another given a specific action. $\mathcal{\gamma} \in \space [0,1]$ is the discount factor, reflecting the agent's trade-off between short-term and long-term interests. The reward function $\mathcal{R}: \mathcal{S}\times\mathcal{A}\times\mathcal{S}\rightarrow\mathcal{R}$ assigns a numerical reward based on the state-action-state transition, providing feedback on the agent's decisions.}

{Within an MDP, at each timestep $t$, the reinforcement learning agent begins by observing the current state $s_t$, then selects an action $a_t$, and finally transitions to a new state $s_{t+1}$ while receiving an immediate reward $r_t$. The resulting sequence of state-action transitions forms a trajectory, and as the agent repeats this process over successive iterations, it collects multiple trajectories of experience. Over time, this continuous interaction enables the agent to learn an optimal policy, denoted by $\pi^*$, that maximizes cumulative rewards. In turn, this optimal policy empowers the agent to accurately identify and extract the key emotional segments embedded within the EEG signals.}

\begin{figure*}[ht]
    \begin{center}
        \includegraphics[width=1.0\textwidth]{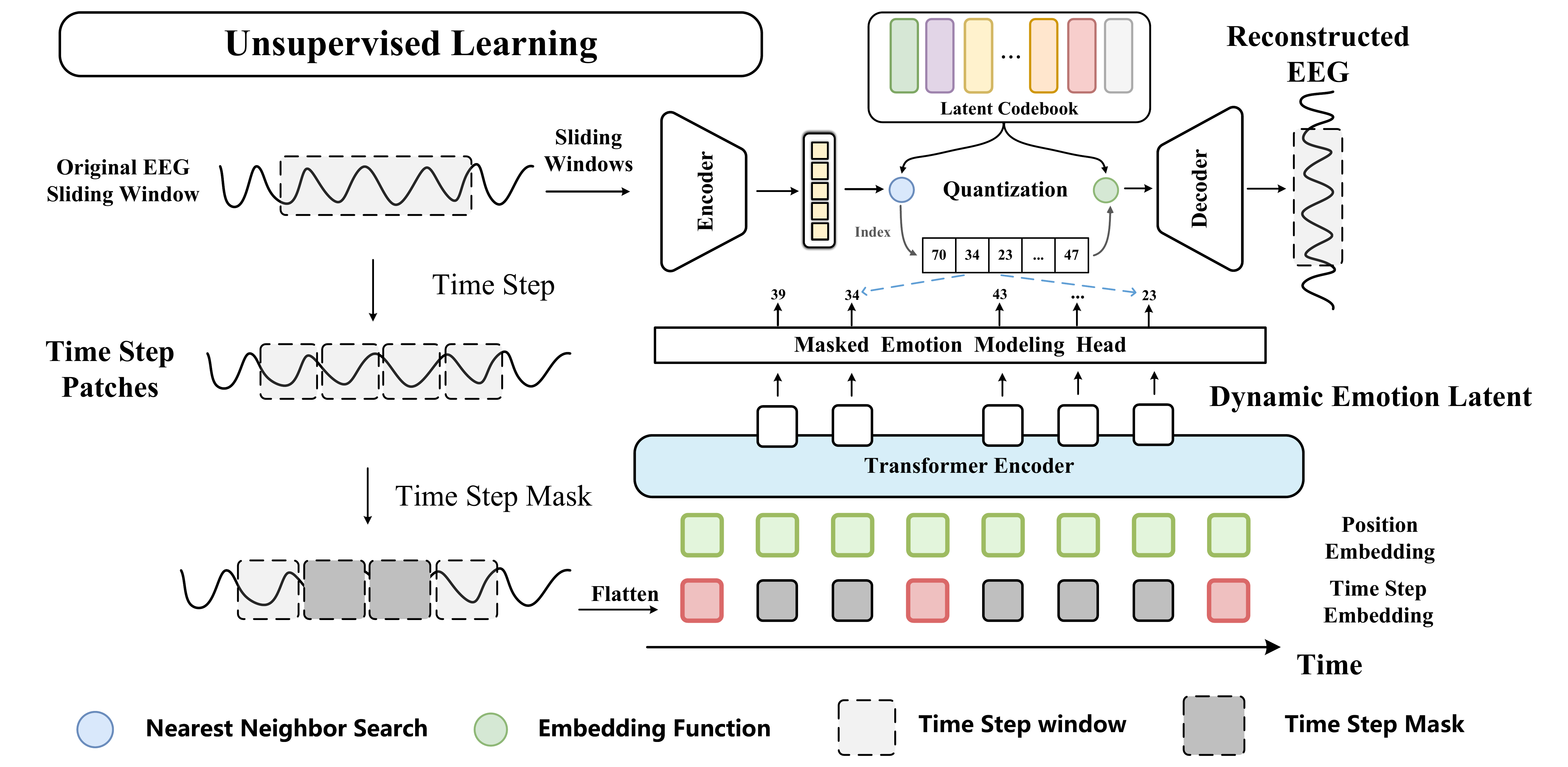}
    \end{center}
    \caption{{Unsupervised Dynamic Emotion Latent Pretraining}}
    \label{fig:Codebook_Transformer_MIM}
\end{figure*}

\begin{figure*}[ht]
    \begin{center}
        \includegraphics[width=1.0\textwidth]{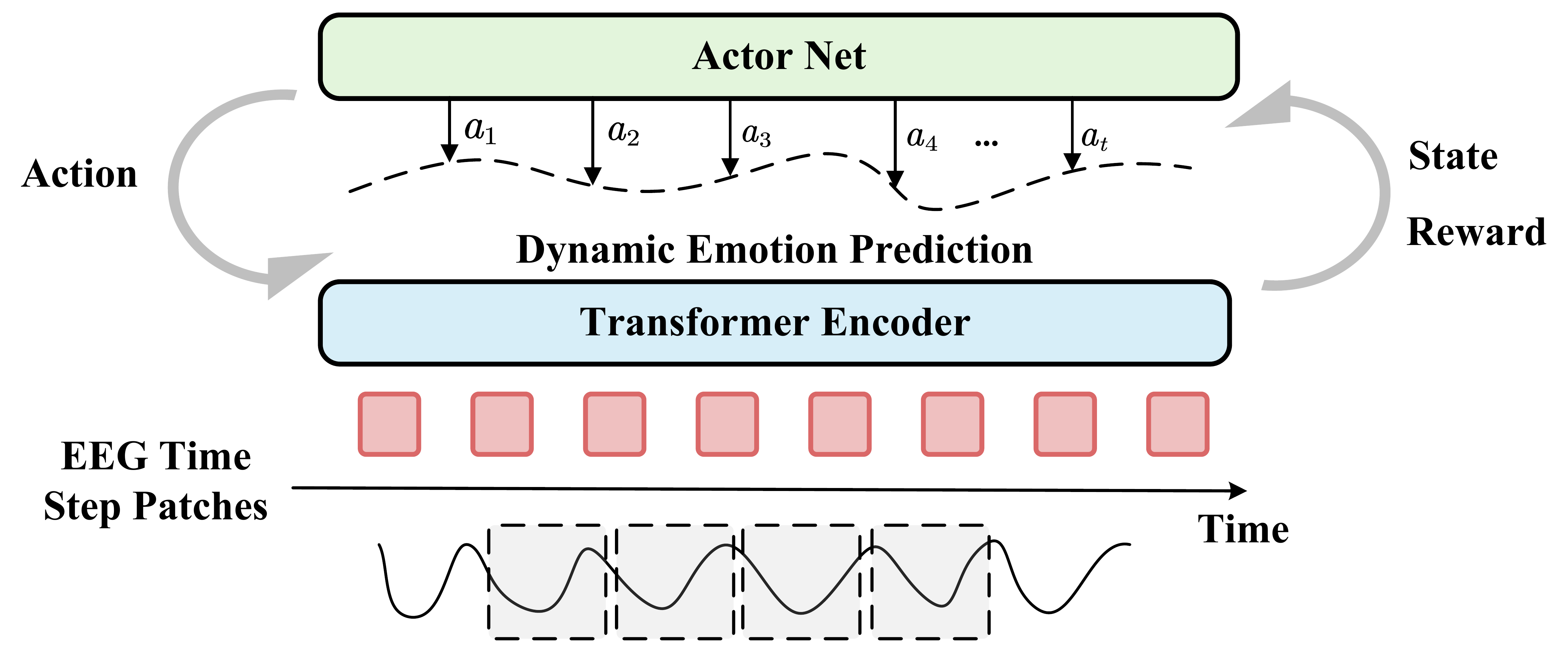}
    \end{center}
    \caption{{The proposed reinforcement Learning Optimization Procss framework.}}
    \label{fig:RL_Optimization}
\end{figure*}

\subsection{{Discerte Codebook Reconstruction}}
\label{subsec:Model Architecture}

{
To enable the model to spontaneously learn the intrinsic data distribution from EEG signal data, we use an autoencoder to reconstruct the EEG data. To learn a more robust feature representation, we introduce VQ-VAE to train the model to develop a discrete, robust representation of the EEG signal's intrinsic nature. Through this VQ-VAE, we can learn a codebook, and any segment of EEG signal can be represented by codebook vectors within this codebook.

To achieve unsupervised extraction of typical emotional states from EEG signals and learning of discrete emotion codebook vector representations, this paper constructs a causal convolutional vector quantization autoencoder\cite{ren2025videoworld}. This model adopts the classic autoencoder paradigm of encoding-quantization-decoding and is specifically designed to address the three major characteristics of EEG signals: temporal causality, spatial topology, and non-stationarity. The core objective is to learn a physiologically meaningful dictionary of typical EEG emotion vectors from massive amounts of unlabeled EEG data, providing a reliable prior for subsequent continuous representation modeling. The model's structure and training process are both based on unsupervised learning, requiring no manually labeled emotion or regression labels.

The VQ-VAE model designed in this study consists of four parts: a spatial encoder, a temporal encoder, a vector quantizer, and a spatiotemporal decoder. The model's input is EEG differential entropy feature data from multiple consecutive time steps. First, the EEG differential entropy features at each time step are treated as an image, and the spatial features at each time step are independently calculated using a convolutional network. Next, the temporal encoder processes the causal temporal relationships at different time steps, and the temporal convolutional network learns the evolutionary relationships of the spatial features at each time step, obtaining the causal spatiotemporal features at each time step. Then, the quantization module quantizes the spatiotemporal features corresponding to different time steps into the nearest codebook vector through nearest neighbor search. Finally, the spatiotemporal decoder uses the corresponding codebook vector to return the original EEG emotion data. The composition of each module of the model will be described below.

\textbf{Spatial Encoder} is built with a convolutional neural network and adopts a two-dimensional convolutional structure. It extracts and reduces the dimensionality of single-time-step EEG spatial features within a sliding window of size $T$. By capturing spatial features at different time steps, the encoder learns spatial correlations between various channels and frequency bands. The convolutional network calculates feature maps of different local regions, and adaptive average pooling compresses spatial features into fixed-dimensional feature vectors. This operation removes redundant spatial information while preserving feature maps that reflect typical spatial structural properties. Finally, fully connected layers project the features into the predefined spatial feature dimension. This procedure maintains the inherent spatial relationships among EEG channels and delivers robust spatial feature representations for subsequent temporal modeling.

\textbf{Temporal Encoder} abandons traditional recurrent neural networks and is constructed based on the Temporal Convolutional Network (TCN). The TCN is capable of capturing the causal dynamic temporal evolution process and processing spatial features of each time step within the sliding window in parallel. Through multi-layer dilated convolution, it expands the temporal receptive field without leaking future information, thereby effectively capturing long-range temporal dependencies. Ultimately, the extraction of spatiotemporal features from EEG signals is accomplished.

\textbf{Vector Quantizer} is the core of the entire VQ-VAE model and the key to learning the codebook vectors of typical emotions in EEG signals, providing the essential codebook prior for this study. The core design of this module is a learnable discrete codebook $E=\{e_1,e_2,\cdots,e_k\}$, where $N$ is the number of codebook vectors. Each codebook vector corresponds to a typical emotional state in the EEG signal, representing the transition process between different emotion categories.\\
The core idea of Vector Quantization (VQ) is to encode arbitrary signals with a group of fixed vectors. These compressed fixed vectors represent the most essential semantic characteristics of signals.Specifically, for the continuous spatiotemporal feature vector \(z_e\) output by the temporal encoder, the squared $L2$ distance between the feature at each individual time step and all codebook vectors is computed. The feature is then mapped to the nearest codebook vector to complete quantization, and the quantized feature \(z_q\) is obtained.To solve the gradient discontinuity problem induced by discrete hard quantization, the Straight-Through Estimator (STE) is employed to realize effective gradient backpropagation. In addition, a codebook reset mechanism is introduced to reduce redundant codebook vectors. Codebook vectors are reinitialized according to the distribution features of active code entries, ensuring favorable utilization efficiency and representation capacity of the codebook.

\textbf{Spatiotemporal decoder} has a symmetrical structure to the encoding part. First, a causal convolutional temporal decoder restores the quantized discrete codebook representation to temporal features. Then, a transposed convolutional layer and a fully connected layer of the spatial decoder map the temporal features back to the spatial dimension of the original EEG signal, ultimately outputting the reconstructed EEG signal $\hat{x}$. The core function of the decoding module is to force the codebook representation learned by the encoding and quantization modules to fully preserve the core information of the original EEG signal through reconstruction constraints, ensuring the effectiveness of the codebook vector.
}

{The training loss function is as follows:
\begin{equation}
\label{equation: VQ-VAE Total Loss function}
    \mathcal{L}_{total} = \mathcal{L}_{recon} + a\cdot\mathcal{L}_{VQ}+\beta\cdot\mathcal{L}_{commit}
\end{equation}

\begin{equation}
\label{equation: VQ-VAE reconstruct loss function}
    \mathcal{L}_{recon} = ||x-\hat{x}||^2
\end{equation}

\begin{equation}
\label{equation: VQ-VAE vector quantization loss function}
    \mathcal{L}_{VQ} = ||sg[Z_e(x)]-z_q^i(x)||^2
\end{equation}

\begin{equation}
\label{equation: VQ-VAE commitment loss function}
    \mathcal{L}_{commit} = ||z_e(x)-sg[z_q^i(x)]||^2
\end{equation}
In eq.(\ref{equation: VQ-VAE Total Loss function}) $\alpha$ and $\beta$ are hyperparameters that balance the weights of different loss terms, and the reconstruction loss is computed using the Mean Squared Error. The first term constrains the error between the reconstructed signal and the input signal, ensuring that the codebook representations fully preserve the core information of the raw input EEG. The second term is the Vector Quantization Loss, which encourages the codebook vectors to fit the encoder output as closely as possible. The third term is the Commitment Loss, which constrains the encoder such that its output approximates the codebook vectors, thereby preventing a disconnect between the encoder output and the codebook.
}

\subsection{{Dynamic Emotion Latent Space Masked Modeling}}
{
\textbf{Motivation}:
{Human emotions are continuously and dynamically changing, not based on discrete emotion modeling; there are transitional processes between different emotions. Therefore, this study further utilizes causal spatiotemporal Codebook for discrete emotion state representation. Through unsupervised masking, the model learns the continuous evolution of dynamic emotions over a long time dimension within the latent space.}

{As stated in the first part and the second part, there are currently many electric signal information, and the information in different areas has been changed, and the model information has been built and the inevitable noise has come. The electrical signal activity reflects the direct physiological signal of the brain's activity, and its analysis, display, mechanical connection, mechanical analysis, etc. are important areas. However, the primitive electric signal usually has non-uniformity, loud noises and different physical differences, and lacks the special ability of the direct acquisition of guidance, and the situation is not fixed under the same scene. Currently based on VQ-VAE, there are many hard-wired methods for displaying the power, and each time the projector can be projected in one code. The ebook book is full of information, but it is actually possible to study separately.
}

Specifically, codebook quantization representation uses discrete hard coding. This method matches each EEG temporal block with only a single codebook vector, failing to depict the continuous changes in emotions and struggling to capture the global correlation patterns of long-term EEG signals. This approach is ill-suited for representing continuous changes in EEG emotions. To address these issues, this study proposes a core modeling approach: assuming the existence of a dynamic emotion latent space, any segment of EEG emotion can be represented as a linearly weighted combination of a fixed set of vectors, plus zero-mean detail residuals. The dynamic emotion latent space is characterized by the following formula.}

\begin{equation}
\label{equation: Dynamic Emotion Latent}
    x=\sum_{k=1}^na_ke_k+\epsilon
\end{equation}
Where, $\{e_k\}_{k=1}^N$ is the codebook vector corresponding to the typical emotional state in the Codebook, $a_k$ is the combination coefficient, which represents the contribution weight of the current EEG emotion by different typical emotions, i.e., the codebook vector, and to a certain extent, it describes the dynamic gradual change of emotion, and $\epsilon$ is the residual of EEG emotion near the typical emotional state.

To realize such a modeling paradigm, this study adopts a Transformer-based unsupervised masking method to predict the semantic information corresponding to typical emotional codebooks. Dynamic emotion modeling is divided into two parts: the generation of discrete tokens for emotional codebooks and unsupervised masked training.First, causal spatial-temporal convolution is used to learn the typical emotional states of EEG signals, and a dictionary is constructed as a prior for spatial-temporal semantics. Then, long-sequence dynamic emotion modeling is conducted through unsupervised masked learning based on the Transformer Encoder. Random masking at the time-step level is performed on the input signals, where the time-step EEG features are zeroed out with a fixed probability to generate the mask. The original signals are projected into the dynamic emotion latent space via a projection layer, and long-sequence dependencies are captured by the self-attention mechanism, so as to predict the spatial-temporal semantic typical emotional states corresponding to the masked time-step EEG emotional features.

The Transformer Encoder captures the long‑term temporal dependencies of EEG emotional features across different time steps within the latent space via the self‑attention mechanism. Through unsupervised masked prediction training, the linear combination coefficients of different typical emotion vectors at each time step are continuously optimized, so as to learn the inherent emotional semantic correlations at each time step. The dynamic emotion representation derived from the unsupervised masking procedure during model training is formulated as follows:
\begin{equation}
\label{equation: unsupervised masked modeling}
    p_{t,k}=Softmax(A_{t,:})_k=\frac{exp(A_{b,t,k})}{\sum_{i=1}^Kexp(A_{b,t,k})}
\end{equation}
Where $p_{t,k}$ denotes the predicted contribution probability of the $k$ codebook at the $t$ time step.

Meanwhile, the codebook vector indices generated by VQ‑VAE for the raw EEG signals serve as the ground-truth labels $y\in\{1,2,...,K\}$.Accordingly, the training loss function is the cross-entropy function, defined as the information entropy difference between the predicted distribution and the true distribution, which is formulated as:
\begin{equation}
\label{equation: cross entropy loss function}
    \mathcal{L}_{CE} = -\frac{1}{T}\sum_{t=1}^T\sum^K_{k=1}q_{t,k}log(p_{t,k})
\end{equation}

By minimizing the cross‑entropy loss, the model is constrained to predict a strict correspondence between the discrete spatiotemporal semantics of the masked features and the discrete spatiotemporal codebook based on contextual information.

Furthermore, to compress the temporal EEG signals into a latent space that is highly relevant to the emotion recognition task, emotion labels are introduced during the masked prediction process to guide the model. This enables the latent space to focus more effectively on task-related dimensions and the information associated with emotional representations. Specifically, the MSE loss of continuous emotion labels is employed as a supervised constraint to encourage the model to learn latent representations that are more relevant to the downstream task. This process also serves as a cold-start stage for subsequent reinforcement learning-based optimization of temporal prediction performance\cite{guo2025deepseek}. The optimization objective is formulated as follows:
\begin{equation}
\label{equation: mean squared error loss function}
    \mathcal{L}_{MSE} = (y_t - \hat{y}_t)^2
\end{equation}
where $y_t$  denotes the ground-truth value, and $\hat{y}_t$ represents the prediction generated by the model.

The optimization objective of the dynamic emotional latent space modeling is formulated as follows:
\begin{equation}
\label{equation: Total obejective of Dynamic Emotion Latent Space}
    \mathcal{L}_{Dynamic} = \mathcal{L}_{CE} + \mathcal{L}_{MSE}
\end{equation}
The optimization objective of the dynamic emotional latent space modeling is formulated through the joint optimization of the cross-entropy loss and the mean squared error loss. On the one hand, the cross entropy loss enhances the model’s capability to capture intrinsic long term temporal dependencies and spatiotemporal pattern combinations. On the other hand, during the process in which the model compresses EEG spatiotemporal features into a low dimensional manifold via an unsupervised masking task, the MSE loss guides the model to focus more on task-relevant feature dimensions during dimensionality reduction. This provides a solid feature representation foundation for subsequent reinforcement learning to further optimize task-oriented objectives.

\subsection{Temporal Prediction Optimization through Reinforcement Learning}
{This paper proposes dynamic emotion modeling based on unsupervised masked dynamic emotion. Subsequently, reinforcement learning, using the dynamic emotion latent space and the subjects' genuine emotional feedback, further optimizes the dynamic emotion latent space, enabling the model to better capture real emotional changes and accurately predict emotions. This section introduces the reward function design and optimization process of the reinforcement learning SAC algorithm during the reinforcement learning phase.}

{
To guide the agent in accurately predicting emotional state scores based on changes in EEG emotional state transitions, and considering our desire for the agent to not only accurately predict state scores but also to understand the trends in predicted score changes during state transitions, we designed three reward functions: a regression prediction reward function, a temporal smooth reward function, and a temporal delta reward function. For trial-level EEG signals, during the agent's sequential decision-making process, the reward functions guide the agent to capture changes in the subject's emotional state while accurately predicting their emotional state score.

\textbf{Regression Prediction Reward} This reward function effectively ensures the accuracy of prediction by calculating the squared error between the continuous predicted value of the agent's continuous emotion at the current time step and the continuous emotion label.
\begin{equation}
\label{Regression Reward Functiuon}
    \mathcal{R}_{reg} = exp(-k_1\cdot(y_t-\hat{y}_t)^2)
\end{equation}
where \(y_t\) denotes the true label at the current time step, and \(\hat{y}_t\) denotes the predicted value of the agent at the current time step.

\textbf{Temporal Smooth Reward} Based on the short-term continuity of human emotions, this is the behavioral constraint for the agent's emotion prediction, which punishes abrupt changes in adjacent prediction values to ensure the continuity of emotion prediction.
\begin{equation}
\label{Temporal smoothing Reward Functiuon}
    \mathcal{R}_{\text {smooth }}= \begin{cases}\exp \left(-k_2 \cdot\left|\hat{y}_t-\hat{y}_{t-1}\right|\right) & t>1 \\ 0 & t=1\end{cases}
\end{equation}
where \(\hat{y}_t\) is the predicted result at the current time step, and \(\hat{y}_{t-1}\) is the predicted result of the agent at the previous time step.

\textbf{Temporal Delta Reward} Based on the long-term similarity of human emotions, we believe that under the same type of emotional stimuli, the overall mean of human emotions will not change significantly, and there is mean regression. The reward function punishes the global average deviation between the emotion prediction result and the actual prediction result, and as a result reward, it reduces the overall deviation of the prediction result.

\begin{equation}
\label{Long Term Similar Reward Function Total}
\mathcal{R}_{delta} = exp(- \left( (\hat{y}_t - \hat{y}_{t-1}) - (y_t - y_{t-1}) \right)^2)
\end{equation}
where $\hat{y}_t$ is the predicted result at the current time step, and $y_t$ is the true continuous emotion label at the current time step.
\begin{equation}
\label{Total Reward}
Reward = \mathcal{R}_{reg} + \mathcal{R}_{smooth} + \mathcal{R}_{delta}
\end{equation}

Guided by the reward function Reward, the agent can capture the transition relationships of brainwave emotional states during the exploration and utilization process, thereby accurately predicting dynamic and continuous emotions.
}

{
\textbf{Reinforcement Learning} We employ the Soft Actor-Critic reinforcement learning algorithm to train the model, predicting the continuous emotional states of subjects based on real-time EEG data at the trial level, and learning the optimal policy through trial and error in the continuous action space. To address the issues of large exploration space and low sample efficiency in the continuous action space, SAC introduces an entropy maximization constraint, enabling the agent to maintain effective exploration in a stochastic dynamic environment, avoiding getting trapped in local optima, thereby improving the model's robustness and generalization ability. SAC is an off-policy algorithm based on the Actor-Critic framework, and its maximum entropy reinforcement learning objective can be expressed as:
\begin{equation}
\label{equation: SAC Optimization Objective}
\pi^*=\arg \max _\pi \mathbb{E}_{\left(\boldsymbol{s}_t, \boldsymbol{a}_t\right)-\rho_\pi}\left\{\sum_t r\left(\boldsymbol{s}_t, \boldsymbol{a}_t\right)+\alpha H\left[\pi\left(\cdot \mid \boldsymbol{s}_t\right)\right]\right\}
\end{equation}
where $\pi$ and $\pi_*$ denote the current policy and the optimal policy, respectively; $r(s_t,a_t)$ represents the immediate reward obtained by the agent; $(s_t,a_t)\sim\rho_{\pi}$ stands for the state-action trajectory distribution induced by policy $\pi; \pi(\cdot \mid s_t)$ is the stochastic policy that maps the current state $s_t$ to a distribution over the action space; $H[\pi(\cdot\mid s_t)]$ denotes the policy entropy, which encourages the agent to explore the state-action space and prevents the policy from converging to local optima; $\alpha$ is the temperature coefficient that balances the weights of the reward and the entropy.
The corresponding Q-function of the soft state-value function is given as follows:
\begin{equation}
\label{equation: SAC Optimization Objective Q value function}
Q\left(s_t, a_t\right)=\gamma \mathbb{E}_{s_{t+1}, a_{t+1}}\left[V\left(s_{t+1}\right)\right]+r\left(s_t, a_t\right)
\end{equation}
where $\gamma$ denotes the reward discount factor; $V(s)$ represents the state-value function, and its computational expression is given as follows.
\begin{equation}
\label{equation: SAC Optimization Objective state value function}
V\left(s_t\right)=\mathbb{E}_{a_t \sim \pi}\left[Q\left(s_t, a_t\right)-\alpha \ln \pi\left(a_t \mid s_t\right)\right]
\end{equation}

Policy update is achieved by minimizing the KL divergence between the policy and the target distribution
\begin{equation}
\label{equation: SAC Optimization Policy Update}
\pi_{\mathrm{new}}=\arg \min _{\pi \in \Pi} D_{\mathrm{KL}}(\pi\left(\cdot \mid s_t\right) \| \frac{\exp \left[\frac{1}{\alpha} Q^\pi \operatorname{old}\left(s_t,\right)\right]}{z^\pi \operatorname{old}\left(s_t\right)})
\end{equation}
where $D_{KL}$ denotes the KL divergence; $Q^{\pi_{old}}(s_t,\cdot)$ is the Q-function under the old policy $\pi_{old}$; and $Z^{\pi_{old}}(s_t)$ is the normalization constant.

During network training, the Critic network fits the value estimation by means of mean squared error (MSE).
\begin{equation}
\label{equation: SAC Optimization Critic Loss function}
J_Q(\theta)=\mathbb{E}_{\left(s_t, a_t\right) \sim D, a_t \sim \pi_\phi}\left[\frac{1}{2}\left[Q_\theta\left(s_t, a_t\right)-r\left(s_t, a_t\right)-\gamma V_{\bar{\theta}}\left(s_{t+1}\right)\right]^2\right]
\end{equation}
where $\theta$ denotes the parameters of the Q-network, $\bar{\theta}$ denotes the parameters of the target Q-network, and $\phi$ denotes the parameters of the policy network; $Q_{\theta}$, $V_{\bar{\theta}}$ and $\pi_{\phi}$ represent the updated functions; $D$ stands for the experience replay buffer.

The objective of the Actor network is:
\begin{equation}
\label{equation: SAC Optimization Actor Optimization Objective}
J(\alpha)=\mathbb{E}_{a_t \sim \pi_t, s_t \sim D}\left[-\alpha \ln \pi\left(a_t \mid s_t\right)-\alpha H_0\right]
\end{equation}
where $H_0$ is the dimension of actions output by the policy network, the Actor network outputs the policy entropy, and $\alpha$ denotes the temperature coefficient.

During training, the model consists of an Actor module and a Critic module. The Actor continuously interacts with the dynamic emotional latent space during decision-making and iteratively searches for the optimal policy based on the reward mechanism. The Critic guides this exploration process by estimating the cumulative expected return for each state-action pair, thereby assisting the Actor in finding the optimal policy.
}

\subsection{{EEG Emotion Keyframe Detection}}
{
This paper achieves continuous emotion prediction using the proposed model and performs accurate emotion recognition based on the prediction results, thereby identifying EEG key frames containing critical emotional moments. The identification of key frames consists of two parts:(1) smoothing the emotion prediction results, and(2) identifying EEG emotional key moments according to the prediction results.
The specific steps are as follows:
(1) Preprocess the emotion label sequence. First, smooth the emotion label sequence obtained through regression to generate a smoothed emotional state curve.
(2) Identify EEG emotional key time points. For the smoothed emotional curve, identify key emotional time points, including extreme points and mutation points. Extreme points correspond to moments when the local emotional intensity reaches the highest or lowest level.
}

\begin{table*}[h]
\caption{Detailed description of the SEED, SEED-IV and Long-Term EEG Emotion Continuous datasets.}
\label{tab:Experiment on datasets}
    \centering
    \scalebox{0.55}{
    \begin{tabular}{@{}c@{\hskip 0.2cm}c@{\hskip 0.2cm}c@{\hskip 0.2cm}c@{\hskip 0.2cm}c@{\hskip 0.2cm}c@{\hskip 0.2cm}c@{\hskip 0.3cm}c@{\hskip 0.3cm}c@{}}
        \toprule
        & Datasets & Subject & Sessions & Trials & Channels & Sampling Rate (Hz) & \multicolumn{2}{c}{Classes}\\
        \midrule
        & SEED & 15 & 3 & 15 & 62 & 1000 & \multicolumn{2}{c}{3 (Negative, Neutral, Positive)}\\ 
        & SEED-IV & 15 & 3 & 24 & 62 & 1000 & \multicolumn{2}{c}{4 (Happy, Neutral, Sad, Fear)}\\ 
        & Long-term Naturalistic Emotion Dataset & 49 & 3 & 12 & 64 & 1000 & 2 Valence & 2 Arousal 2 Dominace\\
        \bottomrule
        \end{tabular}}
\end{table*}

\section{Experiments}
\label{sec:Experiments}
\subsection{{Datasets}}
\label{subsec:Datasets}

{We conduct experiments on three publicly available EEG datasets: SEED \cite{zheng2015investigating}, SEED-IV \cite{zheng2018emotionmeter}, and Private Long-Term Natural Paradigm Emotion Dataset\cite{hu2022microstate,hu2023eeg}. These datasets are widely used benchmarks for evaluating EEG-based emotion recognition algorithms. The dataset statistics are summarized in Table \ref{tab:Experiment on datasets}.}

{\textbf{SEED} EEG signals were recorded from 15 subjects (7 males and 8 females) using a 62-channel ESI NeuroScan system following the international 10-20 standard. The raw EEG data was sampled at 1000 Hz. Each participant completed three experimental sessions. In every session, they viewed 15 film clips designed to elicit emotional responses. These clips were categorized into three emotional states: positive, neutral, and negative, with five clips per category.}

{\textbf{SEED-IV} EEG signals were recorded from 15 subjects (7 males and 8 females) using a 62-channel ESI NeuroScan system along with an SMI eye tracker. Each subject participated in three experimental sessions. In each session, they watched 24 film clips designed to elicit four emotional states: happy, sad, fearful, and neutral. Each emotional state contained six film clips.}

{\textbf{Long-Term Naturalistic Emotion Dataset} The Long-term Naturalistic Emotion Paradigm Dataset was developed by the MINDLAB Laboratory, School of Biomedical Engineering, Shenzhen University. EEG signals were recorded from 49 participants using a 64 channel BrainAmp DC system at a sampling rate of 1 kHz under video-based emotional stimulation.Each participant completed three sessions with identical video materials and presentation order. In total, participants watched 12 video clips (approximately 10 minutes each) designed to induce positive and negative emotions, with six clips for each emotion category.The dataset contains both static and dynamic emotion labels. Static labels describe the overall emotional state of an entire EEG segment as either positive or negative. Dynamic labels were continuously collected during video viewing, with participants providing ratings every 30 seconds on three dimensions: arousal (0–9), valence (0–9), and dominance (0–9).}

{In line with previous research \cite{zhou2025emotion,liu2025eeg}, Differential Entropy (DE) features are extracted from each 1-second EEG segment and used as input to the model.}
{\color{blue}
\begin{table*}[ht]
\centering

\caption{Model Performance for Cross-Subject Emotion Prediction Regression on the SEED Dataset.}
\label{tab:Model performance on SEED}
    \centering
    \resizebox{\textwidth}{!}{
    \begin{tabular}{@{}c@{\hskip 0.2cm}c@{\hskip 0.2cm}c@{\hskip 0.2cm}c@{\hskip 0.2cm}c@{\hskip 0.2cm}c@{\hskip 0.2cm}c@{\hskip 0.2cm}c@{\hskip 0.2cm}c@{}}\\

        \toprule
         Methods & $MSE$ & $MAE$ & Correlation & Methods & $MSE$ & $MAE$ & Correlation \\
        \midrule
        \multicolumn{8}{c}{\textbf{\textit{Traditional Machine Learning Method}}}\\
        \midrule
            \hspace{0.2cm}SVR* \cite{drucker1996support}  & 0.2586 &  0.4033 & 0.4594 &\hspace{0.2cm}KNN* \cite{cover1967nearest} & 0.1749 & 0.3113 & 0.4459\\
            \hspace{0.2cm}Decision Tree* \cite{robinson1965machine}  & 0.2721 &  0.3691 & 0.3042 &\hspace{0.2cm}Random Forest* \cite{breiman2001random} & 0.2721 & 0.3691 & 0.3042\\
            \hspace{0.2cm}Beysian Ridge* \cite{lindley1972bayes}  & 0.2502 &  0.3969 & 0.4494 &\hspace{0.2cm}GradientBoostingRegressor* \cite{friedman2001greedy} & 0.1243 & 0.2832 & 0.6173\\
            \hspace{0.2cm}Lasso* \cite{tibshirani1996regression}  & 0.1281 &  0.2995 & 0.5868 &\hspace{0.2cm}Ridge* \cite{hoerl1970ridge} & 0.1201 & 0.2865 & 0.6367\\
            \hspace{0.2cm}MLP* \cite{rumelhart1986learning}  & 0.1273 &  0.2742 & 0.6113 &\hspace{0.2cm}HuberRegressor* \cite{huber1992robust} & 0.2613 & 0.4048 & 0.4591\\
        \midrule
        \multicolumn{8}{c}{\textbf{\textit{Deep Learning Method}}}\\
        \midrule        
            \hspace{0.2cm}EEGNet* \cite{lawhern2018eegnet}  & 0.1134 &  0.2499 & 0.6628 &\hspace{0.2cm}DAN* \cite{li2018cross} & 0.1770 & 0.3406 & 0.3648\\
            \hspace{0.2cm}DANN* \cite{ganin2016domain}  & 0.1320 &  0.2814 & 0.6634 &\hspace{0.2cm}BiDANN* \cite{li2018novel} & 0.1837 & 0.3710 & 0.3737\\
            \hspace{0.2cm}DDC* \cite{chen2021ms}  & 0.1034 &  0.2469 & 0.6708 &\hspace{0.2cm}DCORAL* \cite{sun2016return} & 0.1201 & 0.2712 & 0.6490\\
            \hspace{0.2cm}RGNN* \cite{zhong2020eeg}  & 0.1656 &  0.3388 & 0.3370 &\hspace{0.2cm}\textbf{Proposed Method} & \textbf{0.0713} & \textbf{0.2012} & \textbf{0.7968}\\

        \bottomrule
        \end{tabular}
        }
\end{table*}

}

\subsection{Implementation Details}
{The overall framework of the emotion prediction model and keyframe recognition based on deep reinforcement learning is mainly divided into three modules: (1) Unsupervised learning of an EEG VQ-VAE based on causal convolution; (2) Pre-training for long-term temporal dependency modeling based on the Transformer; (3) Optimization of emotion prediction based on the SAC reinforcement learning algorithm.

In this study, a causal convolutional Vector-Quantization Variational Autoencoder is employed for feature learning. The model is composed of a spatial encoder, a temporal encoder, a vector quantization layer, a temporal decoder and a spatial decoder. The spatial encoder takes 1-channel spatial features as input, extracts spatial features via two convolutional layers with 32 and 64 channels successively, and finally outputs a 64-dimensional spatial representation. The temporal encoder adopts a 4-layer causal convolutional network structure, with an input dimension of 64, an output dimension of 64, and a convolution kernel size of 3, which is designed to capture temporal dependencies. The vector quantization layer has a commitment loss coefficient of 0.20, and is used to discretize continuous features into discrete codes. The temporal decoder is symmetric to the temporal encoder, and ultimately reconstructs the EEG features with the original spatial dimension.

In the second pre-training stage, a Transformer encoder is adopted to complete the masked modeling task, with a feature dimension of 64, 4 multi-head attention heads, 3 encoder layers, and a maximum sequence length of 128.
For the reinforcement learning-based prediction optimization in the third stage, the pre-trained Transformer encoder from the second stage serves as the feature extractor, and the model structure is kept consistent with that in the pre-training phase. During the reinforcement learning training process, most parameters of the Transformer encoder are frozen, and only a small portion of the encoder parameters are fine-tuned to adapt to the emotion prediction task. This study utilizes the SAC algorithm to implement the continuous emotion value prediction task. The policy network takes the features extracted by the Transformer as input; the value network adopts a dual Q-network structure to mitigate the overestimation problem, takes the concatenated vector of features and 1-dimensional actions as input, and outputs the final Q-value through a fully connected network. The target value network shares the identical structure with the value network and is applied to stabilize the training process.All experiments are implemented based on the PyTorch 1.13.1 framework and conducted on an experimental platform equipped with an NVIDIA GeForce RTX 4090 GPU.}

\begin{table*}[ht]
\centering

\caption{Model Performance for Cross-Subject Emotion Prediction Regression on the SEED-IV Dataset.}
\label{tab:Model performance on SEED-IV}
    \centering
    \resizebox{\textwidth}{!}{
    \begin{tabular}{@{}c@{\hskip 0.2cm}c@{\hskip 0.2cm}c@{\hskip 0.2cm}c@{\hskip 0.2cm}c@{\hskip 0.2cm}c@{\hskip 0.2cm}c@{\hskip 0.2cm}c@{\hskip 0.2cm}c@{}}\\

        \toprule
         Methods & $MSE$ & $MAE$ & Correlation & Methods & $MSE$ & $MAE$ & Correlation \\
        \midrule
        \multicolumn{8}{c}{\textbf{\textit{Traditional Machine Learning Method}}}\\
        \midrule
            \hspace{0.2cm}SVR* \cite{drucker1996support}  & 0.1623 &  0.3238 & 0.3771 &\hspace{0.2cm}KNN* \cite{cover1967nearest} & 0.1918 & 0.3420 & 0.2610\\
            \hspace{0.2cm}Decision Tree* \cite{robinson1965machine}  & 0.1882 &  0.3454 & 0.1538 &\hspace{0.2cm}Random Forest* \cite{breiman2001random} & 0.1086 & 0.2658 & 0.2670\\
            \hspace{0.2cm}Beysian Ridge* \cite{lindley1972bayes}  & 0.1532 &  0.3048 & 0.2932 &\hspace{0.2cm}GradientBoostingRegressor* \cite{friedman2001greedy} & 0.1199 & 0.2756 & 0.2505\\
            \hspace{0.2cm}Lasso* \cite{tibshirani1996regression}  & 0.1064 &  0.2587 & 0.4511 &\hspace{0.2cm}Ridge* \cite{hoerl1970ridge} & 0.1036 & 0.2560 & 0.4316\\
            \hspace{0.2cm}MLP* \cite{rumelhart1986learning}  & 0.1081 &  0.2580 & 0.4697 &\hspace{0.2cm}HuberRegressor* \cite{huber1992robust} & 0.1713 & 0.3327 & 0.3291\\
        \midrule
        \multicolumn{8}{c}{\textbf{\textit{Deep Learning Method}}}\\
        \midrule        
            \hspace{0.2cm}EEGNet* \cite{lawhern2018eegnet}  & 0.1067 &  0.2568 & 0.4408 &\hspace{0.2cm}DAN* \cite{li2018cross} & 0.1075 & 0.2558 & 0.3857\\
            \hspace{0.2cm}DANN* \cite{ganin2016domain}  & 0.1054 &  0.2534 & 0.4089 &\hspace{0.2cm}BiDANN* \cite{li2018novel} & 0.1032 & 0.2492 & 0.4291\\
            \hspace{0.2cm}DDC* \cite{chen2021ms}  & 0.1020 &  0.2480 & 0.4485 &\hspace{0.2cm}DCORAL* \cite{sun2016return} & 0.1046 & 0.2516 & 0.4410\\
            \hspace{0.2cm}RGNN* \cite{zhong2020eeg}  & 0.1154 &  0.2741 & 0.3442 &\hspace{0.2cm}\textbf{Proposed Method} & \textbf{0.0832} & \textbf{0.2036} & \textbf{0.5465}\\

        \bottomrule
        \end{tabular}
        }
\end{table*}

\subsection{{Experimental Results}}
\label{subsec:Experimental Results on the SEED, SEED-IV and DEAP datasets}
{
{To fully verify the effectiveness and accuracy of the proposed model, extensive comparative experiments were carried out on two public datasets and a long-term EEG emotion dataset. The comparison approaches cover traditional machine learning methods and deep learning-based emotion recognition methods.

The traditional machine learning methods include:(1) SVR; (2) KNN; (3) Decision Tree; (4) Random Forest; (5) Bayesian Ridge; (6) Gradient Boosting Regressor; (7) Lasso; (8) Ridge; (9) MLP; (10) Huber Regressor.The deep learning methods include: (1) EEGNet; (2) DAN; (3) DANN; (4) BiDANN; (5) DDC; (6) DCORAL; (7) RGNN.The experimental results comparing the above methods with the method proposed in this study on the SEED, SEED-IV, and long-term natural paradigm emotion datasets are shown below.
}

{Table \ref{tab:Model performance on SEED} presents the regression performance of different models on the SEED dataset under the leave-one-subject-out cross-validation setting. The experimental results demonstrate that the proposed model achieves excellent performance on this dataset for continuous emotion label regression. Specifically, the proposed method achieves a mean squared error (MSE) of 0.0713, a mean absolute error (MAE) of 0.2012, and a Pearson correlation coefficient (PCC) of 0.7968 on the SEED dataset.Compared with the best-performing traditional machine learning method, Ridge, the proposed model reduces the MSE by 0.0488, decreases the MAE by 0.0853, and improves the PCC by 0.1601. Compared with the best-performing deep learning method, DDC, the proposed model reduces the MSE by 0.0321, decreases the MAE by 0.0457, and improves the PCC by 0.1201.

}

{
{Table \ref{tab:Model performance on SEED-IV} presents the regression performance of different models on the SEED-IV dataset under the leave-one-subject-out cross-validation setting. The experimental results demonstrate that the proposed model achieves strong performance on this dataset for continuous emotion label regression. Specifically, the proposed method achieves a MSE of 0.0832, a MAE of 0.2036, and a PCC of 0.5465 on the SEED-IV dataset.Compared with the best-performing traditional machine learning method, Ridge, the proposed model reduces the MSE by 0.0204, decreases the MAE by 0.0524, and improves the PCC by 0.1149. Compared with the best-performing deep learning method, DDC, the proposed model reduces the MSE by 0.0188, decreases the MAE by 0.0444, and improves the PCC by 0.0980.}}

{Table \ref{tab:Model performance on Long-Term Naturalistic Emotion Dataset Arousal} presents the regression performance of different models on the arousal dimension of the Long-term Naturalistic Emotion Paradigm Dataset under the leave-one-subject-out cross-validation setting. The experimental results demonstrate that the proposed model achieves strong performance on this dataset for continuous arousal label regression. Specifically, the proposed method achieves a MSE of 0.4316, a MAE of 0.5443, and a PCC of 0.5979.Compared with the best-performing traditional machine learning method, GradientBoostingRegressor, the proposed model reduces the MSE by 0.1398, decreases the MAE by 0.0878, and improves the PCC by 0.2148. Compared with the best-performing deep learning method, DDC, the proposed model reduces the MSE by 0.0516, decreases the MAE by 0.0476, and improves the PCC by 0.0436.}

{Table \ref{tab:Model performance on Long-Term Naturalistic Emotion Dataset Dominace} presents the regression performance of different models on the dominance dimension of the Long-term Naturalistic Emotion Paradigm Dataset under the leave-one-subject-out cross-validation setting. The experimental results show that the proposed model achieves strong performance for continuous dominance label regression. The proposed method achieves a MSE of 0.2045, a MAE of 0.3107, and a PCC of 0.5794.Compared with the best-performing traditional machine learning method, GradientBoostingRegressor, the proposed model reduces the MSE by 0.0857, decreases the MAE by 0.1327, and improves the PCC by 0.2271. Compared with the best-performing deep learning method, DDC, the proposed model reduces the MSE by 0.0486, decreases the MAE by 0.0820, and improves the PCC by 0.1142.}
}

{Overall, the experimental results on the SEED, SEED-IV dataset and the Long-term Naturalistic Emotion Dataset demonstrate the effectiveness and robustness of the proposed method for continuous emotion regression tasks. The proposed model consistently achieves superior performance across different emotional dimensions, including arousal and dominance, under the leave-one-subject-out cross-validation setting. By effectively capturing discriminative emotional representations from EEG signals and reducing the influence of redundant or noisy information, the proposed method significantly improves regression accuracy and correlation performance compared with both traditional machine learning methods and existing deep learning approaches. These results indicate that the proposed framework has strong generalization ability and provides an effective solution for EEG-based continuous emotion estimation.}

\begin{table*}[ht]
\centering

\caption{Model Performance for Cross-Subject Emotion Prediction Regression on the Long-Term Naturalistic Emotion Dataset of Arousal.}
\label{tab:Model performance on Long-Term Naturalistic Emotion Dataset Arousal}
    \centering
    \resizebox{\textwidth}{!}{
    \begin{tabular}{@{}c@{\hskip 0.2cm}c@{\hskip 0.2cm}c@{\hskip 0.2cm}c@{\hskip 0.2cm}c@{\hskip 0.2cm}c@{\hskip 0.2cm}c@{\hskip 0.2cm}c@{\hskip 0.2cm}c@{}}\\

        \toprule
         Methods & $MSE$ & $MAE$ & Correlation & Methods & $MSE$ & $MAE$ & Correlation \\
        \midrule
        \multicolumn{8}{c}{\textbf{\textit{Traditional Machine Learning Method}}}\\
        \midrule
            \hspace{0.2cm}SVR* \cite{drucker1996support}  & 0.5948 &  0.6124 & 0.2718 &\hspace{0.2cm}KNN* \cite{cover1967nearest} & 0.6832 & 0.6974 & 0.2180\\
            \hspace{0.2cm}Decision Tree* \cite{robinson1965machine}  & 0.6779 &  0.6917 & 0.2362 &\hspace{0.2cm}Random Forest* \cite{breiman2001random} & 0.5845 & 0.6631 & 0.3571\\
            \hspace{0.2cm}Beysian Ridge* \cite{lindley1972bayes}  & 0.6048 &  0.6524 & 0.3558 &\hspace{0.2cm}GradientBoostingRegressor* \cite{friedman2001greedy} & 0.5714 & 0.6321 & 0.3831\\
            \hspace{0.2cm}Lasso* \cite{tibshirani1996regression}  & 0.5855 &  0.6559 & 0.3522 &\hspace{0.2cm}Ridge* \cite{hoerl1970ridge} & 0.6050 & 0.6525 & 0.3556\\
            \hspace{0.2cm}MLP* \cite{rumelhart1986learning}  & 0.5780 &  0.5962 & 0.2645 &\hspace{0.2cm}HuberRegressor* \cite{huber1992robust} & 0.6220 & 0.6546 & 0.3596\\
        \midrule
        \multicolumn{8}{c}{\textbf{\textit{Deep Learning Method}}}\\
        \midrule        
            \hspace{0.2cm}EEGNet* \cite{lawhern2018eegnet}  & 0.5657 &  0.6437 & 0.4206 &\hspace{0.2cm}DAN* \cite{li2018cross} & 0.6039 & 0.6655 & 0.3246\\
            \hspace{0.2cm}DANN* \cite{ganin2016domain}  & 0.5441 &  0.6219 & 0.4543 &\hspace{0.2cm}BiDANN* \cite{li2018novel} & 0.5132 & 0.5919 & 0.4782\\
            \hspace{0.2cm}DDC* \cite{chen2021ms}  & 0.4832 &  0.5919 & 0.5543 &\hspace{0.2cm}DCORAL* \cite{sun2016return} & 0.5476 & 0.6220 & 0.4626\\
            \hspace{0.2cm}RGNN* \cite{zhong2020eeg}  & 0.5832 &  0.6437 & 0.4206 &\hspace{0.2cm}\textbf{Proposed Method}  & \textbf{0.4316} & \textbf{0.5443} & \textbf{0.5979}\\

        \bottomrule
        \end{tabular}
        }
\end{table*}

\begin{table*}[ht]
\centering

\caption{Model Performance for Cross-Subject Emotion Prediction Regression on the Long-Term Naturalistic Emotion Dataset of Dominace.}
\label{tab:Model performance on Long-Term Naturalistic Emotion Dataset Dominace}
    \centering
    \resizebox{\textwidth}{!}{
    \begin{tabular}{@{}c@{\hskip 0.2cm}c@{\hskip 0.2cm}c@{\hskip 0.2cm}c@{\hskip 0.2cm}c@{\hskip 0.2cm}c@{\hskip 0.2cm}c@{\hskip 0.2cm}c@{\hskip 0.2cm}c@{}}\\

        \toprule
         Methods & $MSE$ & $MAE$ & Correlation & Methods & $MSE$ & $MAE$ & Correlation \\
        \midrule
        \multicolumn{8}{c}{\textbf{\textit{Traditional Machine Learning Method}}}\\
        \midrule
            \hspace{0.2cm}SVR* \cite{drucker1996support}  & 0.3540 &  0.4846 & 0.3892 &\hspace{0.2cm}KNN* \cite{cover1967nearest} & 0.3392 & 0.4741 & 0.3939\\
            \hspace{0.2cm}Decision Tree* \cite{robinson1965machine}  & 0.3145 &  0.4575 & 0.3364 &\hspace{0.2cm}Random Forest* \cite{breiman2001random} & 0.2940 & 0.4446 & 0.3892\\
            \hspace{0.2cm}Beysian Ridge* \cite{lindley1972bayes}  & 0.3003 &  0.4484 & 0.3996 &\hspace{0.2cm}GradientBoostingRegressor* \cite{friedman2001greedy} & 0.2902 & 0.4434 & 0.3523\\
            \hspace{0.2cm}Lasso* \cite{tibshirani1996regression}  & 0.2955 &  0.4462 & 0.3355 &\hspace{0.2cm}Ridge* \cite{hoerl1970ridge} & 0.3002 & 0.4483 & 0.4000\\
            \hspace{0.2cm}MLP* \cite{rumelhart1986learning}  & 0.3148 &  0.4575 & 0.3997 &\hspace{0.2cm}HuberRegressor* \cite{huber1992robust} & 0.3035 & 0.4493 & 0.3360\\
        \midrule
        \multicolumn{8}{c}{\textbf{\textit{Deep Learning Method}}}\\
        \midrule        
            \hspace{0.2cm}EEGNet* \cite{lawhern2018eegnet}  & 0.3003 &  0.4484 & 0.3996 &\hspace{0.2cm}DAN* \cite{li2018cross} & 0.2990 & 0.4481 & 0.4724\\
            \hspace{0.2cm}DANN* \cite{ganin2016domain}  & 0.2950 &  0.4444 & 0.4751 &\hspace{0.2cm}BiDANN* \cite{li2018novel} & 0.2800 & 0.4210 & 0.4954\\
            \hspace{0.2cm}DDC* \cite{chen2021ms}  & 0.2531 &  0.3927 & 0.4652 &\hspace{0.2cm}DCORAL* \cite{sun2016return} & 0.2954 & 0.4451 & 0.4893\\
            \hspace{0.2cm}RGNN* \cite{zhong2020eeg}  & 0.3367 &  0.4996 & 0.4931 &\hspace{0.2cm}\textbf{Proposed Method} & \textbf{0.2045} & \textbf{0.3107} & \textbf{0.5794}\\

        \bottomrule
        \end{tabular}
        }
\end{table*}

\begin{table}[ht]
\centering

\caption{Model Performance for Module Ablation Emotion Prediction Regression on the SEED Dataset.}
\label{tab:Module Abalation on SEED}
    \centering
    \begin{tabular}{@{}c@{\hskip 0.6cm}c@{\hskip 0.3cm}c@{\hskip 0.3cm}c@{\hskip 0.3cm}c@{\hskip 0.3cm}c@{}}\\

        \toprule
         Methods & $MSE$ & $MAE$  & $Corr$ & $R^2$\\

        \midrule
        Conv1d + LSTM & 0.1262 & 0.2627 & 0.6538 & 0.4214\\
        Transformer & 0.1218 & 0.2539 & 0.6923 & 0.4531\\
        Ours Without Transformer and Mask & 0.1088  & 0.2401 & 0.6895 & 0.4946\\
        Ours Without VQ-VAE & 0.0993  & 0.2318 & 0.7257 & 0.5134\\
        Ours Supervised Learning Without RL & 0.0852 & 0.2232 & 0.7538 & 0.5519\\
        \textbf{Proposed Method} & \textbf{0.0713} & \textbf{0.2012} & \textbf{0.7968} & \textbf{0.5881} \\
        

        \bottomrule
        \end{tabular}

\end{table}

\begin{table}[ht]
\centering

\caption{Model Performance for Module Ablation Emotion Prediction Regression on the SEED-IV Dataset.}
\label{tab:Module Abalation on SEED-IV}
    \centering
    \begin{tabular}{@{}c@{\hskip 0.6cm}c@{\hskip 0.3cm}c@{\hskip 0.3cm}c@{\hskip 0.3cm}c@{\hskip 0.3cm}c@{}}\\

        \toprule
         Methods & $MSE$ & $MAE$  & $Corr$ & $R^2$\\

        \midrule
        Conv1d + LSTM & 0.0995 & 0.2374 & 0.4579 & 0.2234\\
        Transformer & 0.0997 & 0.2268 & 0.4719 & 0.2295\\
        Ours Without Transformer and Mask & 0.1028  & 0.2326 & 0.4485 & 0.2206\\
        Ours Without VQ-VAE & 0.0972  & 0.2297 & 0.4792 & 0.2317\\
        Ours Supervised Learning Without RL & 0.0939 & 0.2185 & 0.5063 & 0.2432\\
        \textbf{Proposed Method} & \textbf{0.0832} & \textbf{0.2036} & \textbf{0.5465} & \textbf{0.2874} \\
        

        \bottomrule
        \end{tabular}

\end{table}

\begin{table}[ht]
\centering

\caption{Model Performance for Module Ablation Emotion Prediction Regression on the Long-Term Naturalistic Emotion Dataset of Arousal.}
\label{tab:Module Abalation on Long-Term Continuous dataset of Arousal}
    \centering
    \begin{tabular}{@{}c@{\hskip 0.6cm}c@{\hskip 0.3cm}c@{\hskip 0.3cm}c@{\hskip 0.3cm}c@{\hskip 0.3cm}c@{}}\\

        \toprule
         Methods & $MSE$ & $MAE$  & $Corr$ & $R^2$\\

        \midrule
        Conv1d + LSTM & 0.4945 & 0.6043 & 0.4425 & 0.1242\\
        Transformer & 0.4817 & 0.5912 & 0.4820 & 0.1326\\
        Ours Without Transformer and Mask & 0.4935  & 0.6036 & 0.5062 & 0.1573\\
        Ours Without VQ-VAE & 0.4693  & 0.5684 & 0.5437 & 0.1734\\
        Ours Supervised Learning Without RL & 0.4439 & 0.5632 & 0.5624 & 0.1835\\
        \textbf{Proposed Method} & \textbf{0.4316} & \textbf{0.5443} & \textbf{0.5979} & \textbf{0.2058} \\
        

        \bottomrule
        \end{tabular}

\end{table}

\begin{table}[ht]
\centering

\caption{Model Performance for Module Ablation Emotion Prediction Regression on the Long-Term Naturalistic Emotion Dataset of Dominace.}
\label{tab:Module Abalation on Long-Term Continuous dataset of Dominace.}
    \centering
    \begin{tabular}{@{}c@{\hskip 0.6cm}c@{\hskip 0.3cm}c@{\hskip 0.3cm}c@{\hskip 0.3cm}c@{\hskip 0.3cm}c@{}}\\

        \toprule
         Methods & $MSE$ & $MAE$  & $Corr$ & $R^2$\\

        \midrule
        Conv1d + LSTM & 0.2761 & 0.4042 & 0.4954 & 0.0951\\
        Transformer & 0.2558 & 0.3851 & 0.5127 & 0.1178\\
        Ours Without Transformer and Mask & 0.2425  & 0.3550 & 0.5123 & 0.1236\\
        Ours Without VQ-VAE & 0.2319  & 0.3374 & 0.5269 & 0.1245\\
        Ours Supervised Learning Without RL & 0.2246 & 0.3217 & 0.5461 & 0.1573\\
        \textbf{Proposed Method} & \textbf{0.2045} & \textbf{0.3107} & \textbf{0.5794} & \textbf{0.1732} \\
        

        \bottomrule
        \end{tabular}

\end{table}

\begin{table}[ht]
\centering

\caption{Model Performance for Reward Abalation Emotion Regression on the SEED Dataset.}
\label{tab:Reward Abalation on SEED}
    \centering
    
    \begin{tabular}{@{}c@{\hskip 0.2cm}c@{\hskip 0.2cm}c@{\hskip 0.2cm}c@{\hskip 0.4cm}c@{\hskip 0.4cm}c@{\hskip 0.4cm}c@{}}\\

        \toprule
         $R_{reg}$ & $R_{smooth}$ & $R_{delta}$ &  $MSE$ & $MAE$ & $Corr$ & $R^2$ \\

        \midrule
        $\checkmark$ &  &   & 0.0816 & 0.2114 & 0.7720 & 0.5732 \\
        $\checkmark$ & $\checkmark$ &   & \textbf{0.0713} & \textbf{0.2012} & \textbf{0.7968} & \textbf{0.5881} \\
        $\checkmark$ &  & $\checkmark$  & 0.0842 & 0.2210 & 0.7632 & 0.5693 \\

        & $\checkmark$ & $\checkmark$  & 0.0785 & 0.2094 & 0.7761 & 0.5773 \\
        $\checkmark$  & $\checkmark$ & $\checkmark$  & 0.0736 & 0.2081 & 0.7834 & 0.5842 \\

        \bottomrule
        \end{tabular}
        
\end{table}

\begin{table}[ht]
\centering

\caption{Model Performance for Reward Abalation Emotion Regression on the SEED-IV Dataset.}
\label{tab:Reward Abalation on SEED-IV}
    \centering
    
    \begin{tabular}{@{}c@{\hskip 0.2cm}c@{\hskip 0.2cm}c@{\hskip 0.2cm}c@{\hskip 0.4cm}c@{\hskip 0.4cm}c@{\hskip 0.4cm}c@{}}\\

        \toprule
         $R_{reg}$ & $R_{smooth}$ & $R_{delta}$ &  $MSE$ & $MAE$ & $Corr$ & $R^2$ \\

        \midrule
        $\checkmark$ &  &   & 0.0912 & 0.2124 & 0.5236 & 0.2631 \\
        $\checkmark$ & $\checkmark$ &   & \textbf{0.0832} & \textbf{0.2036} & \textbf{0.5465} & \textbf{0.2874}\\
        $\checkmark$ &  & $\checkmark$  & 0.0948 & 0.2145 & 0.5184 & 0.2536\\

         & $\checkmark$ & $\checkmark$  & 0.0890 & 0.2089 & 0.5264 & 0.2662  \\
        $\checkmark$  & $\checkmark$ & $\checkmark$   & 0.0876 & 0.2032 & 0.5396 & 0.2745 \\

        \bottomrule
        \end{tabular}
        
\end{table}

\begin{table}[ht]
\centering

\caption{Model Performance for Reward Abalation Emotion Regression on the Long-Term Naturalistic Emotion Dataset of Arousal.}
\label{tab:Reward Abalation on Long-Term Naturalistic Emotion Dataset of Arousal}
    \centering
    
    \begin{tabular}{@{}c@{\hskip 0.2cm}c@{\hskip 0.2cm}c@{\hskip 0.2cm}c@{\hskip 0.4cm}c@{\hskip 0.4cm}c@{\hskip 0.4cm}c@{}}\\

        \toprule
         $R_{reg}$ & $R_{smooth}$ & $R_{delta}$ &  $MSE$ & $MAE$ & $Corr$ & $R^2$ \\

        \midrule
        $\checkmark$ &  &   & 0.4378 & 0.5528 & 0.5771 & 0.1962 \\
        $\checkmark$ & $\checkmark$ &   & \textbf{0.4316} & \textbf{0.5443} & \textbf{0.5979} & \textbf{0.2058} \\
        $\checkmark$ &  & $\checkmark$ & 0.4437 & 0.5572 & 0.5436 & 0.1835  \\

        & $\checkmark$ & $\checkmark$  & 0.4416 & 0.5564 & 0.5625 & 0.1972 \\
        $\checkmark$  & $\checkmark$ & $\checkmark$  & 0.4362 & 0.5486 & 0.5871 & 0.2034 \\

        \bottomrule
        \end{tabular}
        
\end{table}

\begin{table}[ht]
\centering

\caption{Model Performance for Reward Abalation Emotion Regression on the Long-Term Naturalistic Emotion Dataset of Dominace.}
\label{tab:Reward Abalation on Long-Term Naturalistic Emotion Dataset of Dominace}
    \centering
    
    \begin{tabular}{@{}c@{\hskip 0.2cm}c@{\hskip 0.2cm}c@{\hskip 0.2cm}c@{\hskip 0.4cm}c@{\hskip 0.4cm}c@{\hskip 0.4cm}c@{}}\\

        \toprule
         $R_{reg}$ & $R_{smooth}$ & $R_{delta}$ &  $MSE$ & $MAE$ & $Corr$ & $R^2$ \\

        \midrule
        $\checkmark$ &  &   & 0.2128 & 0.3154 & 0.5586 & 0.1464 \\
        $\checkmark$ & $\checkmark$ &   & \textbf{0.2045} & \textbf{0.3107} & \textbf{0.5795} & \textbf{0.1732}\\
        $\checkmark$ &  & $\checkmark$  & 0.2286 & 0.3286 & 0.5639 & 0.1525 \\

        & $\checkmark$ & $\checkmark$ & 0.2174 & 0.3173 & 0.5694 & 0.1518  \\
        $\checkmark$  & $\checkmark$ & $\checkmark$   & 0.2156 & 0.3149 & 0.5671 & 0.1673 \\

        \bottomrule
        \end{tabular}
        
\end{table}

\section{Discussion}
\label{subsec:Discussion}{
\subsection{{Module Ablation Study}}
{
{To comprehensively validate the dynamic characteristics of the proposed deep reinforcement learning-based emotion prediction model with a dynamic emotional latent space, ablation studies were conducted on the SEED, SEED-IV, and Long-Term Naturalistic Emotion datasets in this section, as shown in \Cref{tab:Module Abalation on SEED,tab:Module Abalation on SEED-IV,tab:Module Abalation on Long-Term Continuous dataset of Arousal,tab:Module Abalation on Long-Term Continuous dataset of Dominace.}. In these experiments, three modules were separately ablated: (1) the causal convolution-based VQ-VAE; (2) the Transformer-based unsupervised masked dynamic emotion modeling module; and (3) the SAC reinforcement learning policy optimization module. In addition, the 1D-CNN + LSTM and Transformer models were introduced as baseline models for comparison, in order to further verify the effectiveness of each module in the proposed framework.}

{The VQ-VAE maps EEG signals into a structured discrete emotional latent space by performing Vector Quantization on EEG features to learn a representative codebook, thereby achieving the disentanglement of core semantic emotional features. Meanwhile, the causal convolution design strictly guarantees temporal causality in sequence modeling and captures contextual information of dynamic emotional evolution. Experimental results show that after removing the VQ-VAE module, the MSE increased to 0.0993 and the Correlation decreased to 0.7257 on the SEED dataset; on the SEED-IV dataset, the MSE increased to 0.0972 and the Correlation decreased to 0.4792; on the Long-Term Naturalistic Emotion Dataset Arousal, the MSE increased to 0.4693 and the Correlation decreased to 0.5437; and on the Long-Term Naturalistic Emotion Dataset (Dominance), the MSE increased to 0.2319 and the Correlation decreased to 0.5269. These results further demonstrate that removing the VQ-VAE leads to the loss of the core capabilities of noise filtering and semantic disentanglement. In addition, without the structured discrete latent space constraints learned through the Vector Quantization codebook, the subsequent Transformer latent space degenerates into a noisy high-dimensional continuous state, making it difficult to effectively learn universal patterns of emotional dynamic evolution, thereby reducing prediction accuracy.

Based on the pretrained spatiotemporal codebook, this component employs a Transformer with unsupervised masking to model long-term emotional dynamics. In the latent space, EEG emotions are compressed into a continuous temporal manifold, enabling the model to better capture contextual relationships in EEG emotional signals. Experimental results show that after removing the Transformer module, both regression accuracy and the ability to track dynamic emotional changes declined. On the SEED dataset, the MSE increased to 0.1088 and the Correlation decreased to 0.6895; on the SEED-IV dataset, the MSE increased to 0.0972 and the Correlation decreased to 0.4792. On the Long-Term Naturalistic Emotion Dataset Arousal, the MSE increased to 0.4935 and the Correlation decreased to 0.5062; on the Long-Term Naturalistic Emotion Dataset Dominance, the MSE increased to 0.2425 and the Correlation decreased to 0.5123. These findings highlight that the dynamic emotional latent space formed through the unsupervised masking process is beneficial for capturing long-term temporal dependencies in EEG emotions. Furthermore, unsupervised learning enables the model to better capture the intrinsic continuous dynamical relationships of EEG emotional states, thereby improving its ability to predict emotional variation trends.

This component formulates EEG-based temporal prediction as a Markov Decision Process, enabling accurate prediction of emotional states based on EEG emotional context. After removing reinforcement learning and replacing it with direct point-to-point supervised fitting of labeled data, the MSE on the SEED dataset increased to 0.0852 and the Correlation decreased to 0.7538; on the SEED-IV dataset, the MSE increased to 0.0939 and the Correlation decreased to 0.5063; on the Long-Term Naturalistic Emotion Dataset of Arousal, the MSE increased to 0.4439 and the Correlation decreased to 0.5624; and on the Long-Term Naturalistic Emotion Dataset of Dominance, the MSE increased to 0.2246 and the Correlation decreased to 0.5461. The experimental results indicate that supervised learning optimizes only single-timestep losses and focuses on local optimality at the current frame, while ignoring the long-term temporal continuity of human emotions. By introducing the SAC reinforcement learning algorithm with a carefully designed reward function, this study guides the model to learn globally optimal prediction strategies that conform to emotional dynamic evolution patterns, rather than merely pursuing local single-step optimality. As a result, the model achieves higher prediction accuracy, better predicts real-time human emotional states, and more effectively tracks continuous emotional dynamics.}

{In summary, the experiments demonstrate that the VQ-VAE learns a discrete codebook through Vector Quantization, thereby discretizing the latent space. Under the clustering-structured discrete constraints, the Transformer compresses EEG emotional representations into a continuous temporal manifold in the latent space, making EEG emotional changes predictable. This further enhances the model capability to capture temporal dependencies and predict emotional variation trends in EEG signals, enabling accurate emotion prediction and effective tracking of dynamic emotional changes. Moreover, the learned latent representations provide environmental priors for the subsequent reinforcement learning agent, allowing the model to converge more rapidly.}}

\subsection{{Reward Ablation Study}}
{
To comprehensively validate the individual effects of the three reward functions proposed in the reinforcement learning stage, extensive ablation experiments were further conducted on the SEED, SEED-IV, and Long-Term Naturalistic Emotion datasets, as shown in \cref{tab:Reward Abalation on SEED,tab:Reward Abalation on SEED-IV,tab:Reward Abalation on Long-Term Naturalistic Emotion Dataset of Arousal,tab:Reward Abalation on Long-Term Naturalistic Emotion Dataset of Dominace}. In these experiments, the three reward functions were combined in pairs under the following settings: (1) using only the Regression Prediction reward; (2) using both Regression Prediction and Temporal Smooth rewards; (3) using both Regression Prediction and Temporal Delta rewards; (4) using both Temporal Smooth and Temporal Delta rewards; and (5) simultaneously using all three reward functions.

The first major conclusion is that reinforcement learning can further enhance sequential prediction performance even when the optimization objective is consistent with that of supervised learning. As shown in Table \ref{tab:Model performance on SEED}, the MSE decreased to 0.0816, with a reduction of 0.0035, while the Correlation increased to 0.7720, improving by 0.0182. In Table \ref{tab:Module Abalation on SEED-IV}, the MSE decreased to 0.0912, with a reduction of 0.0027, and the Correlation increased to 0.5236, improving by 0.0173. In Table \ref{tab:Reward Abalation on Long-Term Naturalistic Emotion Dataset of Arousal}, the MSE decreased to 0.4378, with a reduction of 0.0061, while the Correlation increased to 0.5771, improving by 0.0147. In Table \ref{tab:Reward Abalation on Long-Term Naturalistic Emotion Dataset of Dominace}, the MSE decreased to 0.2128, with a reduction of 0.0118, and the Correlation increased to 0.5486, improving by 0.0125.

Overall, compared with the traditional supervised learning paradigm based on point-to-point sample mapping, reinforcement learning is more effective in modeling the continuous dynamic evolution process of EEG emotions. Conventional supervised learning can be regarded as a form of local sample-level optimization, which mainly focuses on fitting the relationship between individual time steps or local segments and their corresponding labels. Consequently, it often ignores the continuous temporal evolution patterns of emotional states and is therefore more likely to become trapped in local optima. In contrast, reinforcement learning formulates the sequential emotion prediction task as a Markov Decision Process and performs trajectory-level global optimization by modeling and optimizing the entire sequential prediction process as a unified whole. This enables the model to more comprehensively capture the dynamic characteristics of EEG emotional states and learn their intrinsic temporal continuity patterns, ultimately achieving superior overall prediction performance.

The second major conclusion concerns the design of reward functions. Excessive supervised rewards may constrain the exploration capability of reinforcement learning and reduce the upper bound of potential performance improvement. As shown in \Cref{tab:Reward Abalation on SEED,tab:Reward Abalation on SEED-IV,tab:Reward Abalation on Long-Term Naturalistic Emotion Dataset of Arousal,tab:Reward Abalation on Long-Term Naturalistic Emotion Dataset of Dominace}, the second reward setting, which combines the Regression Prediction reward with the Temporal Smooth reward, consistently achieved SOTA performance. Notably, the Temporal Smooth reward is an unsupervised reward function.

Specifically, in Table \ref{tab:Reward Abalation on SEED}, using only the second reward setting achieved a further reduction of 0.0023 in MSE and an increase of 0.0134 in Correlation compared with using all three reward functions simultaneously. Moreover, compared with the combination of two supervised rewards, namely Regression Prediction and Temporal Delta, the second reward setting further reduced the MSE by 0.014 and improved the Correlation by 0.0188. In Table \ref{tab:Reward Abalation on SEED-IV}, the second reward setting reduced the MSE by 0.0044 and improved the Correlation by 0.0069 compared with the three-reward setting, while further reducing the MSE by 0.0116 and increasing the Correlation by 0.0338 compared with the two-supervised-reward setting. Similarly, in Table \ref{tab:Reward Abalation on Long-Term Naturalistic Emotion Dataset of Arousal}, the second reward setting reduced the MSE by 0.0046 and improved the Correlation by 0.0024 compared with the three-reward setting, while further reducing the MSE by 0.0121 and increasing the Correlation by 0.0336 compared with the two-supervised-reward setting. In Table \ref{tab:Reward Abalation on Long-Term Naturalistic Emotion Dataset of Dominace}, the second reward setting reduced the MSE by 0.0111 and improved the Correlation by 0.0059 compared with the three-reward setting, while further reducing the MSE by 0.013 and increasing the Correlation by 0.0241 compared with the two-supervised-reward setting.

These experimental results demonstrate that excessive reliance on manually annotated supervisory signals during the reinforcement learning exploration process may limit the exploration space of the model and consequently reduce its performance upper bound. In contrast, combining a moderate amount of supervised rewards with unsupervised prior-guided rewards can potentially produce better results. On the one hand, supervised rewards guide reinforcement learning toward solutions that are closer to real-world targets. On the other hand, unsupervised rewards encourage the exploration of latent solutions that may not explicitly appear in the annotated training data. The combination of these two mechanisms further enhances the generalization capability of the model after reinforcement learning optimization.
}

\begin{figure}[h]
    \begin{center}
        \includegraphics[width=1.0\textwidth]{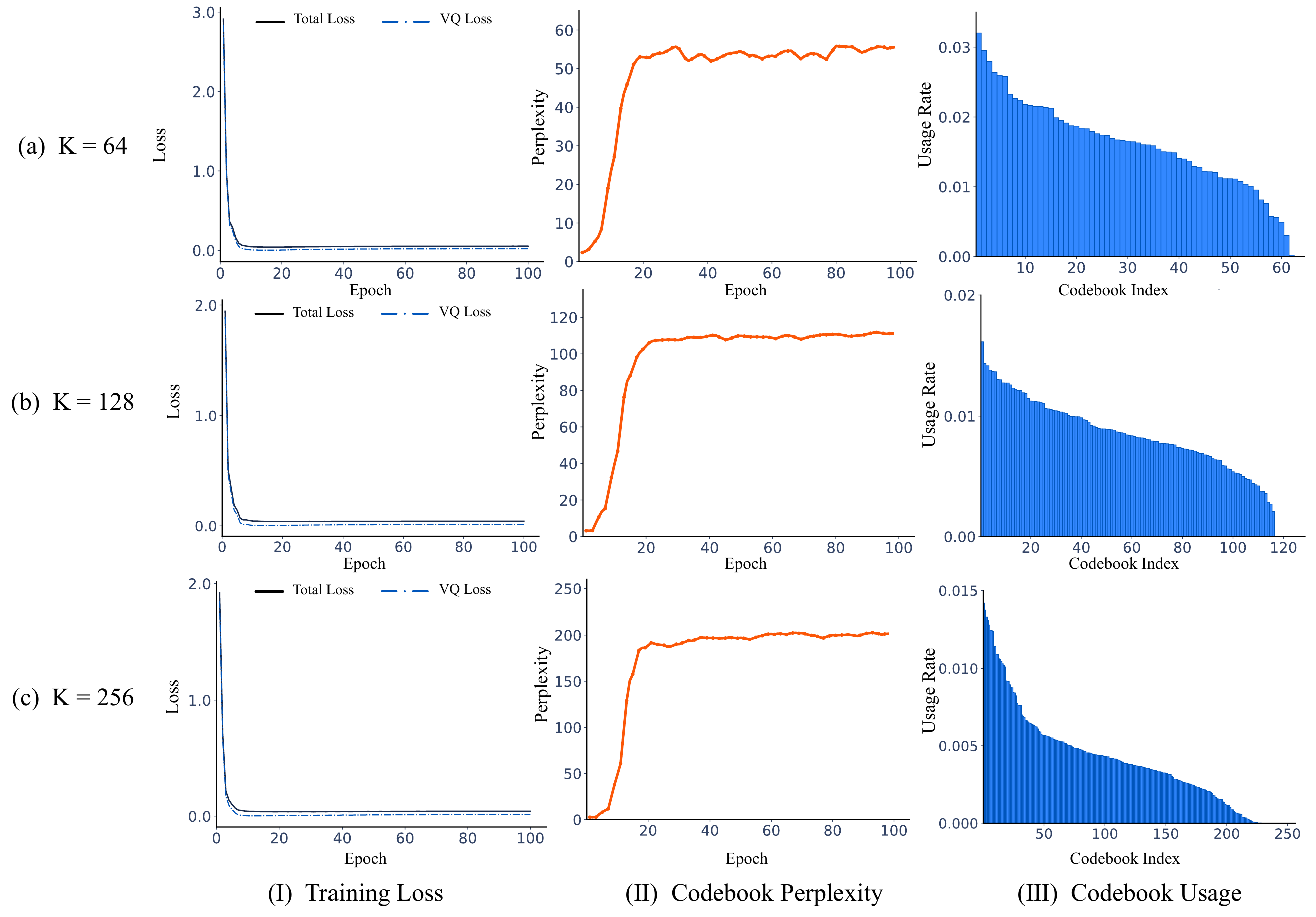}
    \end{center}
    \caption{Visualization Codebook Training Process}
    \label{fig:Visualization Codebook Training Process}
\end{figure}

\begin{figure}[h]
    \begin{center}
        \includegraphics[width=1.0\textwidth]{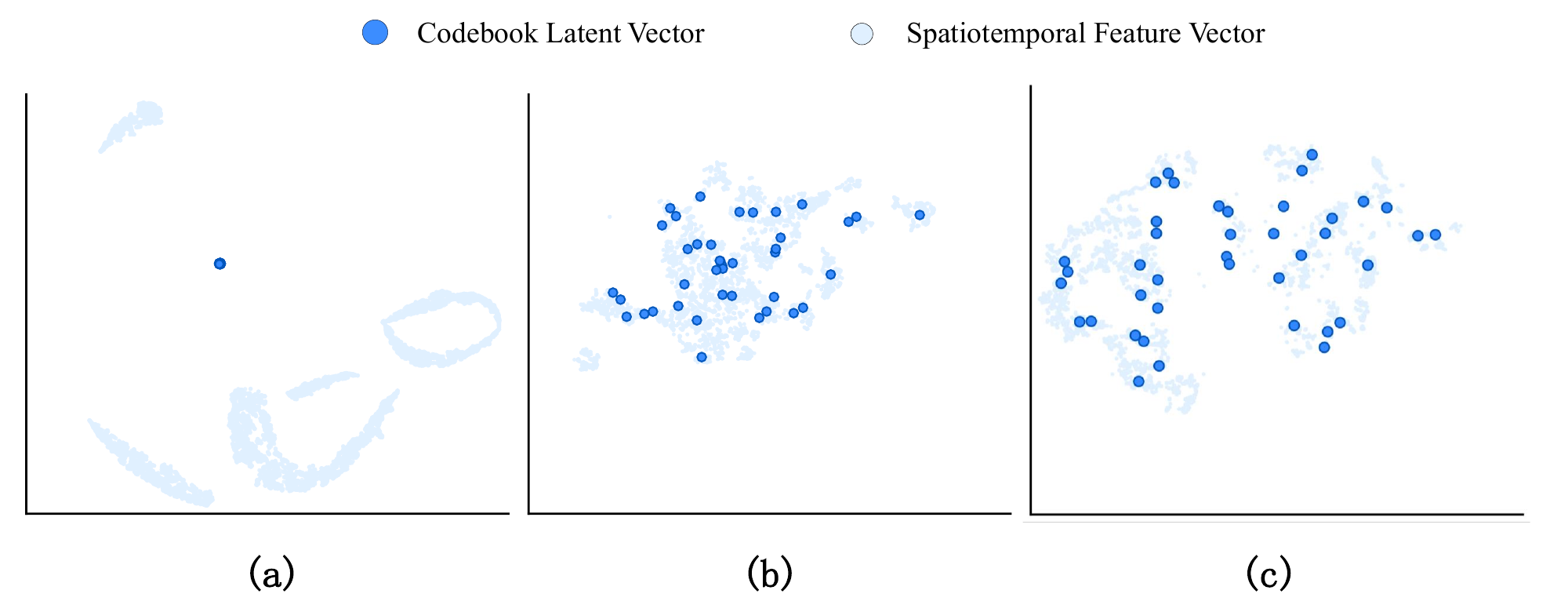}
    \end{center}
    \caption{Visualization t-SNE Codebook Vector}
    \label{fig:Visualization t-SNE Codebook vector}
\end{figure}

\begin{figure}[h]
    \begin{center}
        \includegraphics[width=1.0\textwidth]{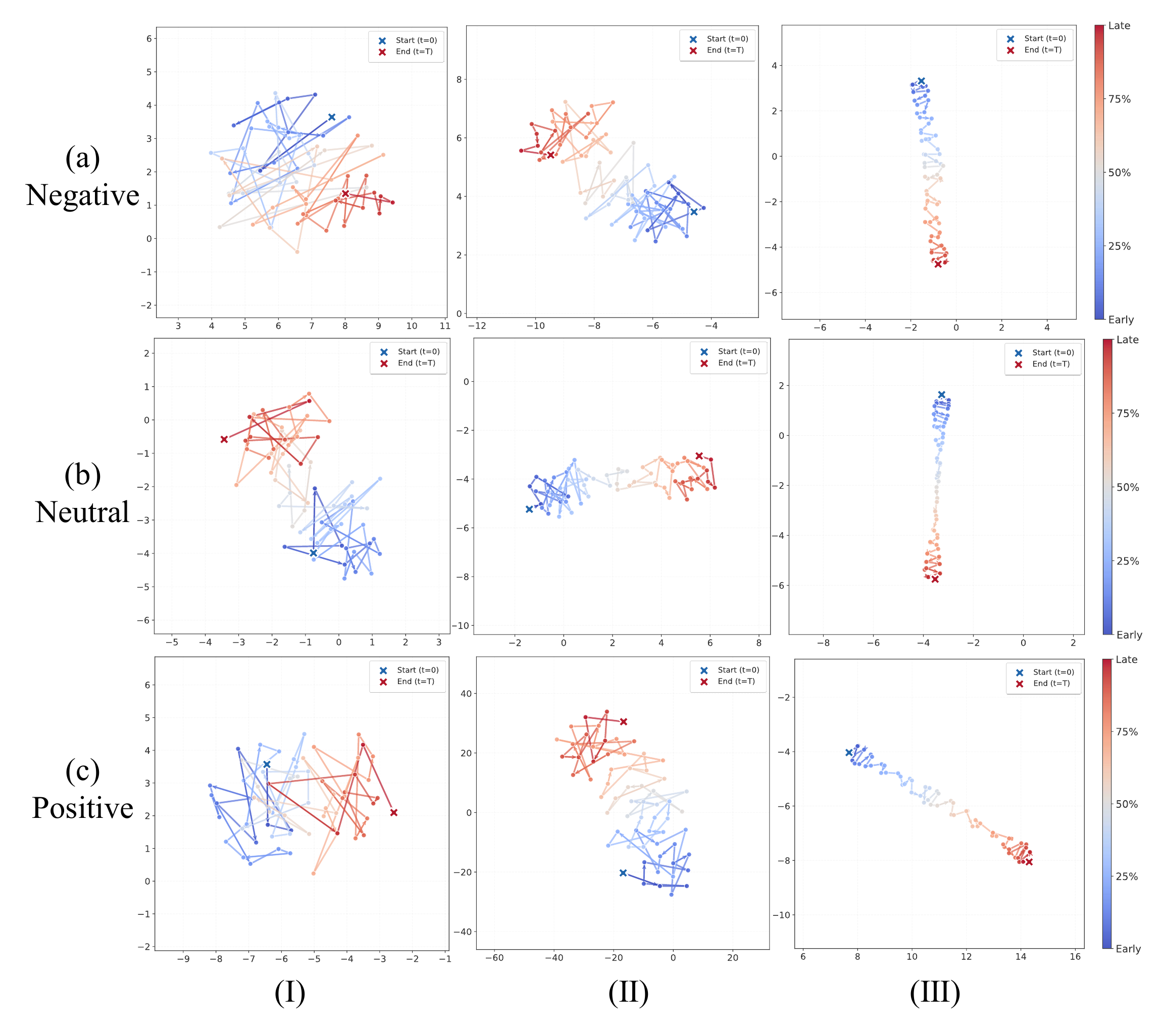}
    \end{center}
    \caption{Comparative visualization of changes in diverse emotions within the dynamic emotional latent space throughout training. Rows from top to bottom correspond to the training processes of (a) negative emotion, (b) neutral emotion and (c) happy emotion, while columns from left to right represent three training stages: (\uppercase\expandafter{\romannumeral 1}) early training stage, (\uppercase\expandafter{\romannumeral 2}) mid-training stage and (\uppercase\expandafter{\romannumeral 3}) training completion.}
    \label{fig:Dynamic Emotion Latent Trajectory}
\end{figure}

\begin{figure}[h]
    \begin{center}
        \includegraphics[width=1.0\textwidth]{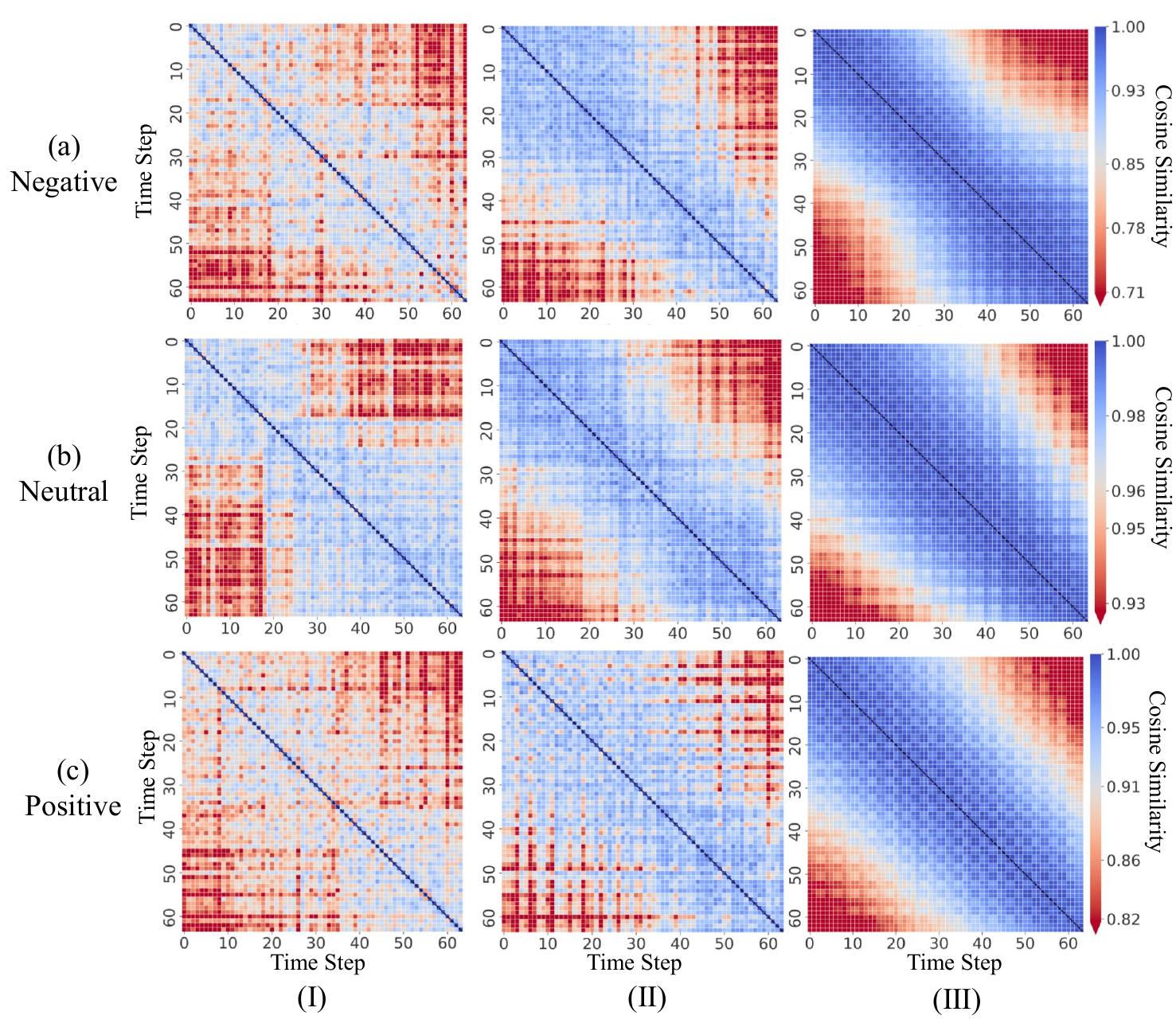}
    \end{center}
    \caption{Heatmaps of feature similarity across different emotion categories in the dynamic emotion latent space during the training process. Rows from top to bottom correspond to the training procedures of three emotion types: (a) negative emotion, (b) neutral emotion, (c) positive emotion. Columns from left to right represent three training phases: (\uppercase\expandafter{\romannumeral 1}) early training stage, (\uppercase\expandafter{\romannumeral 2}) middle training stage, (\uppercase\expandafter{\romannumeral 3}) completed training.}
    \label{fig:Dynamic Emotion Latent Cosine Similarity}
\end{figure}

\begin{figure}[h]
    \begin{center}
        \includegraphics[width=1.0\textwidth]{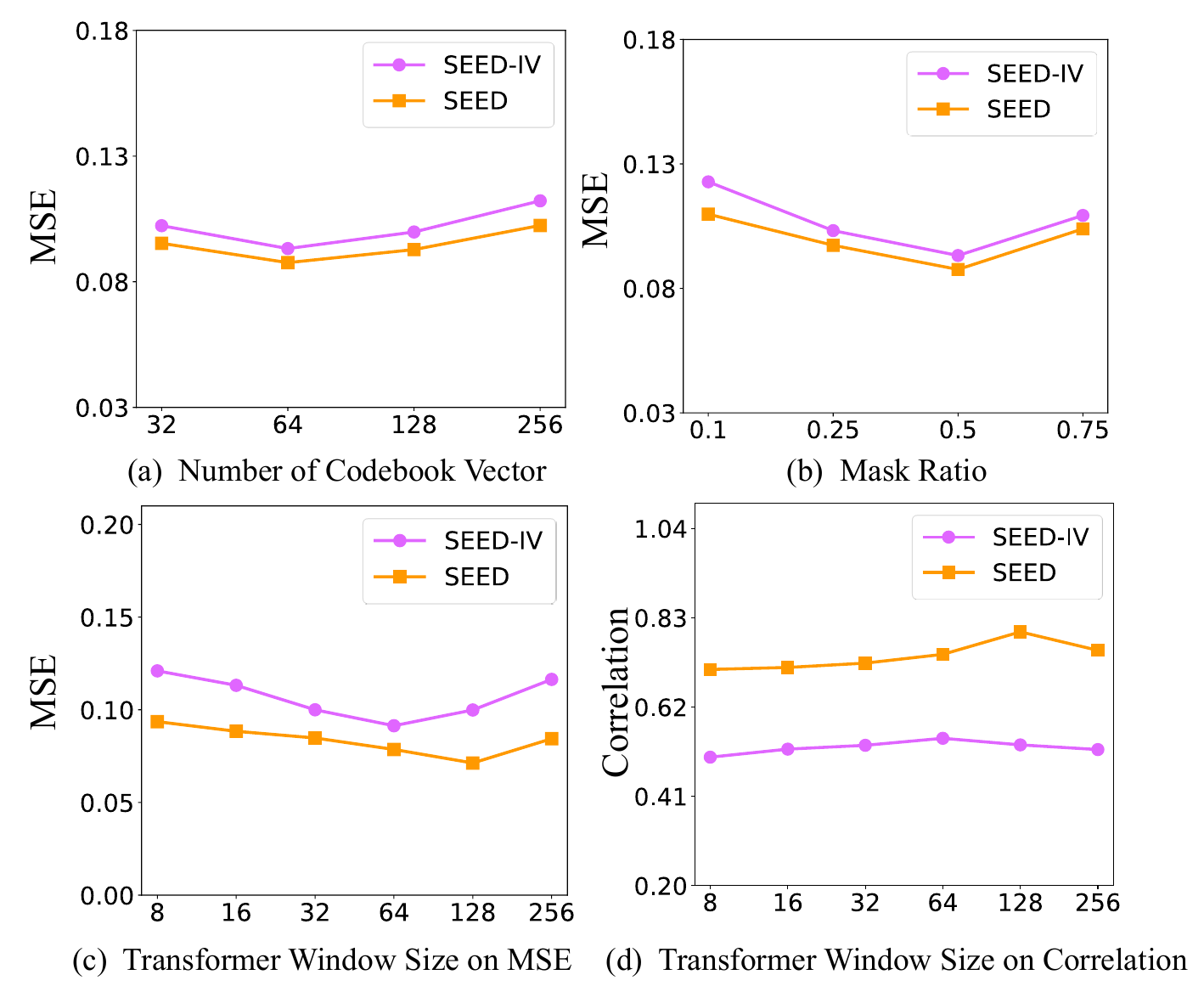}
    \end{center}
    \caption{Sensitivity analysis of hyperparameters.Best
viewed in color. Zoom in for a better view}
    \label{fig:Sensitivity analysis of hyperparameters}
\end{figure}
}

\begin{figure}[h]
    \begin{center}
        \includegraphics[width=1.0\textwidth]{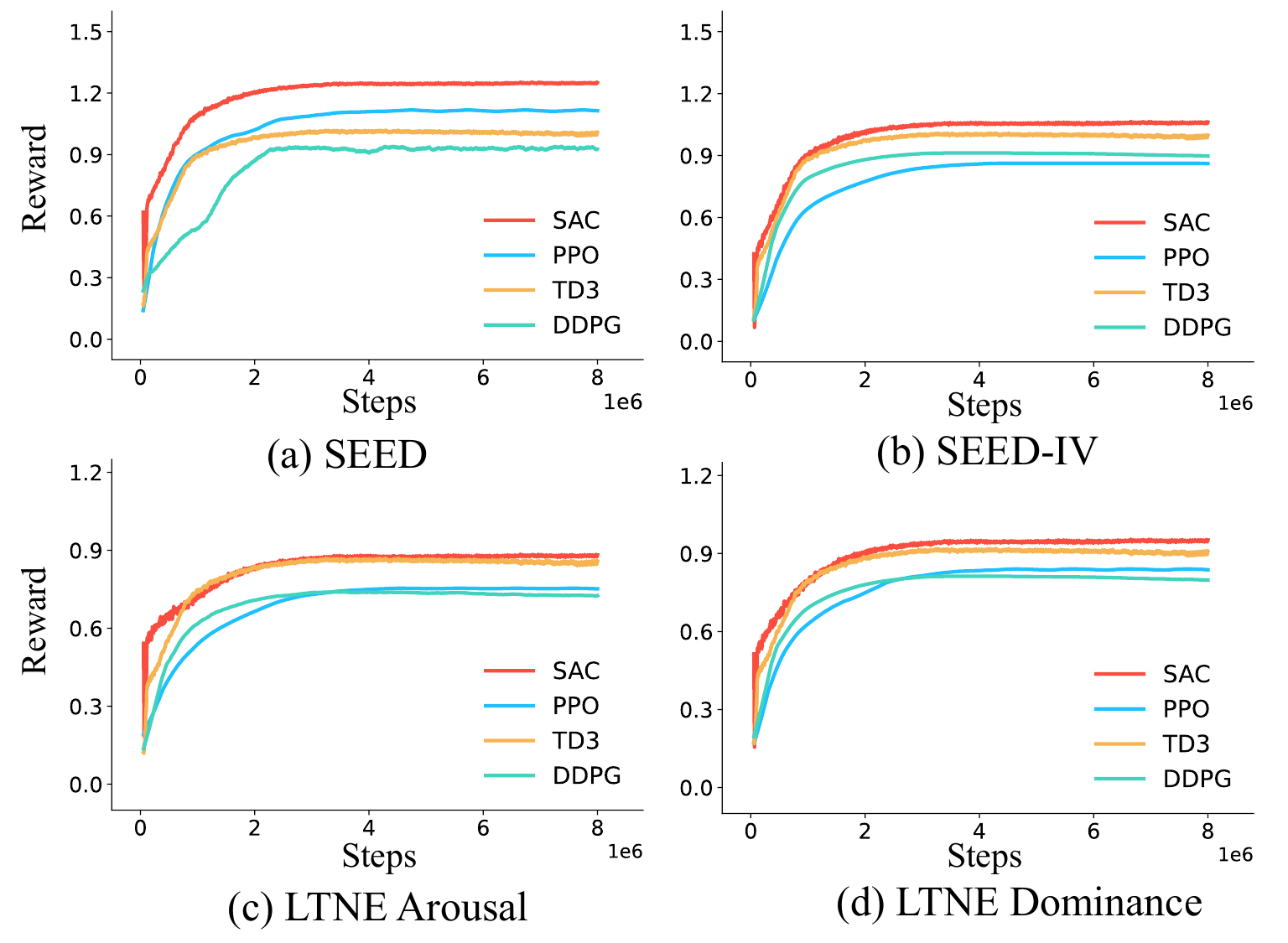}
    \end{center}
    \caption{Reinforcement learning Training Curves on Continuous Regression.}
    \label{fig:RL Algorithm Reward Curvel}
\end{figure}

\begin{figure}[h]
    \begin{center}
        \includegraphics[width=1.0\textwidth]{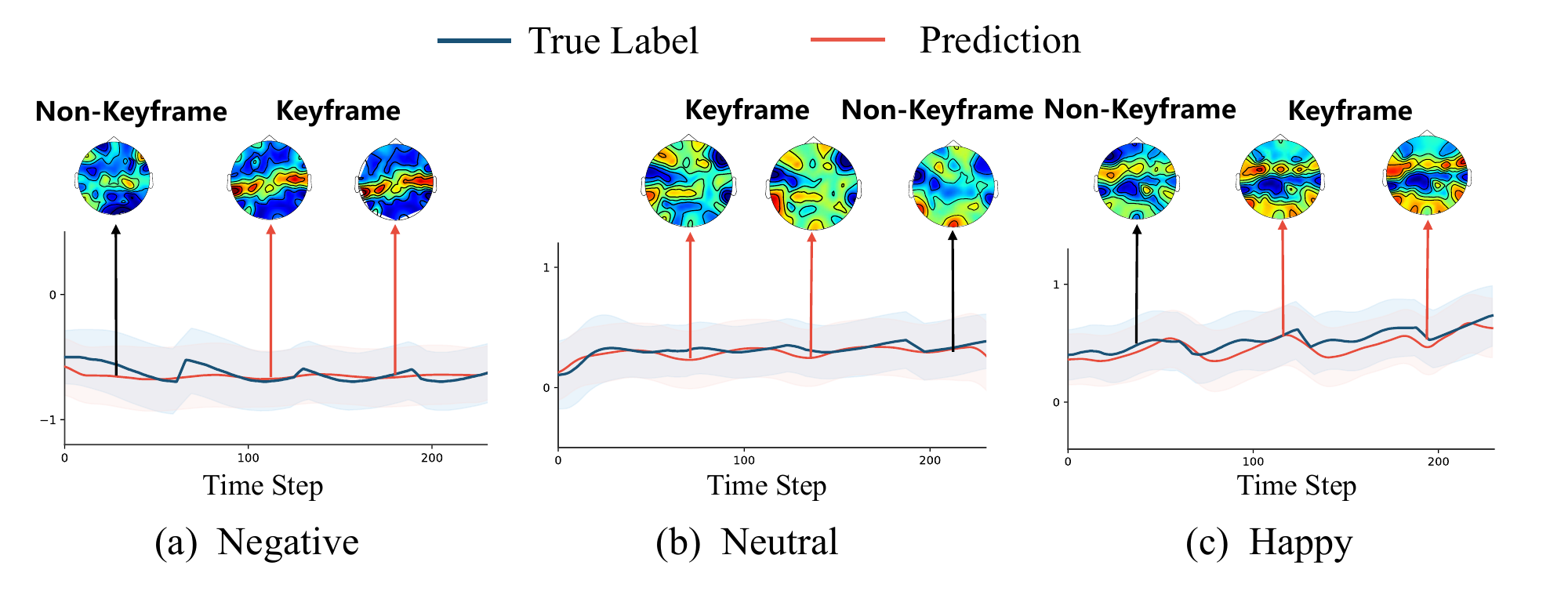}
    \end{center}
    \caption{Evaluation on Keyframe Vislization of Emotion Prediction and the Brain Visualization}
    \label{fig:Keyframe detection}
\end{figure}
}

\subsection{{Codebook Training Analysis}}
{
We first evaluated the training stability and codebook utilization of the proposed vector-quantized autoencoder for EEG signals. As illustrated in Fig.\ref{fig:Visualization Codebook Training Process}, the evaluation includes the training loss curves, codebook perplexity, and the utilization statistics of the learned codebook vectors.

As shown in Fig.\ref{fig:Visualization Codebook Training Process} (\uppercase\expandafter{\romannumeral 1}), the proposed autoencoder employs straight-through gradient estimation, enabling both the reconstruction loss and quantization loss to decrease steadily and converge stably throughout training. Figure\ref{fig:Visualization Codebook Training Process} (\uppercase\expandafter{\romannumeral 2}) further demonstrates that the incorporation of codebook frequency resetting and Exponential Moving Average updates effectively improves the codebook learning process. Regardless of the codebook size, these mechanisms significantly enhance the perplexity and utilization frequency of the codebook vectors. Furthermore, the visualizations in Fig.\ref{fig:Visualization Codebook Training Process} (\uppercase\expandafter{\romannumeral 3}) show that all codebook vectors are actively utilized during quantization, without the occurrence of severely underutilized codebook entries. These results indicate that the proposed method achieves stable optimization and efficient codebook representation learning.

To further investigate the representation learning capability of the proposed causal spatiotemporal EEG vector-quantized autoencoder, we visualized the evolution of the latent-space vectors and codebook vectors during training using t-SNE, as shown in Figure  \ref{fig:Visualization t-SNE Codebook vector}. During the early training stage shown in Fig.\ref{fig:Visualization t-SNE Codebook vector}(a), both the light blue points latent space vectors  and dark blue points codebook vectors are widely scattered, exhibiting no obvious clustering structure. This suggests that the latent representations are still randomly distributed and the quantization process has not yet formed meaningful semantic organization.In the middle stage of training, illustrated in Fig.\ref{fig:Visualization t-SNE Codebook vector}(b), the distance between latent-space vectors and codebook vectors gradually decreases, while several local regions begin to form relatively compact clusters. This phenomenon indicates that the latent representations are progressively aligned with the learned codebook vectors, suggesting that the quantization process is becoming increasingly optimized, although the global clustering structure remains indistinct.After training converges, as shown in Fig.\ref{fig:Visualization t-SNE Codebook vector}(c), the latent-space distribution exhibits a clear and compact clustering structure. The intra-cluster distances are significantly reduced, and each cluster becomes more concentrated. These clusters correspond to representative emotional states embedded in the EEG signals, demonstrating that the proposed model successfully captures the intrinsic distribution characteristics and semantic structure of emotional EEG representations.
}

\subsection{{Dynamic Emotion Latent Space Modeling Analysis}}
{This study proposes a dynamic emotion modeling algorithm based on unsupervised masking. By analyzing the latent space training process from two different perspectives (1) the changes in T-SNE distributions of latent space features during training, and (2) the changes in cosine similarity heatmaps between latent-space features at different time steps during training the study further reveals how dynamic emotions of different categories are modeled throughout the training process.

Figure \ref{fig:Dynamic Emotion Latent Trajectory} illustrates the changes in the time-ordered distributions of EEG data from different emotions in the latent space during the unsupervised masked training process of the Transformer. Through further analysis, this study explains the process of dynamic emotion modeling. From top to bottom, the figure presents different emotion categories, namely sadness, neutrality, and happiness. From left to right, it shows different training stages: the early stage, middle stage, and final stage of training.

For the sadness emotion, as shown in Fig.\ref{fig:Dynamic Emotion Latent Trajectory}(a), a latent-space feature trajectory of a segment of sad EEG emotion data is presented. The starting point in time is represented in blue, while the ending point is represented in red. The EEG emotion latent-space trajectory demonstrates that the model compresses EEG emotions into a continuous temporal manifold. Temporal changes highly correspond to variations in EEG emotional features, and the directions of EEG emotional changes are highly consistent and predictable. This indicates that, within the dynamic emotion latent space, emotional representations are no longer merely discrete latent-space representations, but also characterize the continuous dynamic evolution process of EEG emotions.

During training, the model discretizes the latent space through a causal spatiotemporal codebook, forming a clustering-structured latent space. Based on this structured codebook prior, the Transformer performs unsupervised masked training and predicts the semantic information of masked EEG features through contextual reasoning. EEG emotions with discrete clustering structures are compressed into a continuous temporal manifold, making emotional transitions predictable. This further enhances the model’s ability to capture long-term temporal dependencies in EEG emotions. The model learns the intrinsic continuous evolution patterns of emotional states while also providing environmental priors for subsequent reinforcement learning agents.

As shown in Fig.\ref{fig:Dynamic Emotion Latent Trajectory}(b), a latent-space feature trajectory of neutral emotion EEG data is presented. The neutral emotion latent-space trajectory exhibits similar characteristics. In the early training stage, highly discrete clustering structures are clearly visible. As masked training progresses, EEG emotional features are compressed into a continuous temporal manifold, and the dynamic transition processes between different emotional states are clearly characterized. At the completion of training, EEG features corresponding to different time points in neutral emotions become relatively more compact, which is consistent with the stable and low-fluctuation characteristics of human neutral emotions. This demonstrates that the model has successfully captured the temporal evolution patterns of neutral emotions in EEG signals.

As shown in Fig.\ref{fig:Dynamic Emotion Latent Trajectory}(c), a latent-space feature trajectory of happy emotion EEG data is presented. The happy emotion EEG feature trajectories exhibit similar dynamic characteristics. In the early stage of training, the separation between starting points is significant. As training progresses, the transition processes between different emotional states are compressed into a temporal manifold, showing a more linear structure in which emotional state transitions become predictable. Compared with other emotions, the latent-space trajectories of happy emotion EEG features display larger fluctuations and broader movement ranges, which is consistent with the greater variability typically observed in human happiness. This indicates that the model successfully captures the temporal evolution patterns of happy emotions in EEG signals.

The sample trajectories of happy emotions exhibit consistent evolutionary directions, with extremely high separability between early and late representation spaces and strong clustering properties among samples of the same emotion. The model successfully captures the unique temporal dynamics of happy emotion EEG signals. The trajectories of happy emotion samples exhibit highly linear horizontal temporal paths, where all samples evolve from the blue starting points on the far left toward the right along the horizontal axis, eventually converging at the red endpoints on the far right. The color gradients perfectly correspond to temporal progression, and trajectories of samples belonging to the same emotion category highly overlap.

In summary, within the dynamic emotion latent space, EEG emotion representations are not only discrete latent-space representations of different emotional categories, but also continuous dynamic evolution processes. The combination of discrete emotional semantics and continuous temporal evolution provides a more comprehensive representation of the characteristics of EEG emotions.
}

{
Figure \ref{fig:Dynamic Emotion Latent Cosine Similarity} illustrates the changes in cosine similarity between latent-space features at different time steps for EEG data of different emotions during the unsupervised masked training process of the Transformer. The cosine similarities between different time steps are presented in the form of heatmaps. Through further analysis, this study explains the process of dynamic emotion modeling. From top to bottom, the figure presents different emotion categories, namely sadness, neutrality, and happiness. From left to right, it shows different training stages: the early stage, middle stage, and final stage of training.

First, the changes among different emotion categories are analyzed from top to bottom. For the sadness emotion, as shown in the first row of Figure.\ref{fig:Dynamic Emotion Latent Cosine Similarity} (a), blue indicates higher similarity, while red indicates larger differences. It can be observed that the temporal dependencies between different time steps are more localized. In the heatmap, the blue regions are more concentrated near the diagonal or within small local blocks, indicating that the sadness emotion model focuses more on local temporal dependencies.

For the neutral emotion, as shown in the second row of Figure \ref{fig:Dynamic Emotion Latent Cosine Similarity} (b), the temporal dependency variations are smoother and more stable. The blue bands near the diagonal in the heatmap are more uniform, indicating that the model can more easily capture such temporal dependency relationships.

For the happy emotion, as shown in the third row of Figure \ref{fig:Dynamic Emotion Latent Cosine Similarity} (c), the temporal dependency changes occur at a faster rhythm, causing representations at non-adjacent time steps to also exhibit high similarity. This reflects the stronger fluctuations and more dynamic temporal dependency relationships of happy emotions.

In summary, the model completes the learning of different emotions through the unsupervised masked training process. Distinct differences exist among different emotional categories, while sequences belonging to the same emotion learn specific and structured temporal correlation representations.
}

\subsection{{Evaluation on Reinforcment Learning Optimization}}
\label{subsec:Evaluation on Reinforcment Learning Optimization}{

{To comprehensively validate the superiority of the Soft Actor-Critic (SAC) algorithm in continuous action space optimization, this study conducts extensive comparative experiments among four representative reinforcement learning algorithms: (1) SAC\cite{haarnoja2018soft}, (2) Proximal Policy Optimization\cite{schulman2017proximal} (PPO), (3) Twin Delayed Deep Deterministic Policy Gradient\cite{fujimoto2018addressing} (TD3), and (4) Deep Deterministic Policy Gradient\cite{silver2014deterministic} (DDPG). Comprehensive experiments are performed on three public datasets, including SEED, SEED-IV, and Long-Term Naturalistic Emotion (LTNE), as illustrated in the figure \ref{fig:RL Algorithm Reward Curvel}.

Experimental results demonstrate that the SAC-based reinforcement learning framework consistently achieves SOTA performance across four regression tasks on the three datasets. By introducing the maximum entropy reinforcement learning mechanism, SAC encourages the agent to sufficiently explore the continuous action space, thereby improving the generalization capability of the model. Meanwhile, the adoption of a dual-Critic architecture effectively alleviates the Q-value overestimation problem, further enhancing the temporal prediction performance.As illustrated in the reward curves, both TD3 and DDPG belong to the class of Deterministic Policy Gradient methods, exhibiting relatively small oscillations during the early training stage and demonstrating favorable convergence stability. In particular, TD3 introduces twin Q-networks and delayed policy update mechanisms, which mitigate the Q-value overestimation issue existing in DDPG to a certain extent. Consequently, TD3 achieves better overall performance and training stability than DDPG. However, since both TD3 and DDPG adopt deterministic policies, their exploration capabilities in continuous action spaces remain limited, making them prone to falling into local optima. As a result, the reward growth becomes slower in the later training stage, and their final performance remains significantly inferior to SAC.In contrast, PPO, as a representative Policy Gradient method, improves training stability by constraining the policy update range. Nevertheless, its relatively conservative policy update strategy limits effective exploration in high-dimensional continuous action spaces for complex temporal prediction tasks, resulting in lower final rewards compared to SAC. Particularly on challenging emotional dynamic modeling tasks such as the SEED dataset, PPO exhibits a noticeable slowdown in convergence during the later training phase, indicating certain limitations in modeling long-term temporal dependencies.Further analysis of the experimental results reveals that SAC not only rapidly improves the reward value during the early training stage but also maintains stable growth in the later stage, ultimately achieving the highest cumulative rewards. This indicates that SAC can more effectively balance the trade-off between exploration and exploitation, thereby possessing stronger policy optimization capability in complex non-stationary temporal environments. In addition, the maximum entropy mechanism introduces higher stochasticity and robustness into the policy, enhancing the model’s adaptability to different subjects, emotional states, and cross-scenario data distribution shifts.Moreover, on the Arousal and Dominance tasks of the LTNE dataset, the performance gap among the reinforcement learning algorithms becomes even more pronounced. This observation suggests that as the complexity of temporal dependencies increases, the exploration capability and long-term value estimation ability of reinforcement learning algorithms directly influence the final model performance. Benefiting from the Soft Bellman Backup mechanism and stochastic policy optimization strategy, SAC can estimate long-term cumulative returns more accurately, thereby exhibiting stronger advantages in complex emotional temporal prediction tasks.

In summary, the experimental results fully demonstrate the effectiveness and superiority of the SAC algorithm in continuous action space temporal prediction tasks. SAC consistently outperforms traditional reinforcement learning methods such as PPO, TD3, and DDPG in terms of convergence speed, training stability, exploration capability, and final prediction performance, thereby validating the feasibility and superiority of the proposed framework for affective computing and EEG temporal modeling tasks.
}
}

{
\subsection{Visualization of Emotion Prediction and Keyframe Brain Topology}

{This study proposes a mask-based long-sequence dynamic emotion modeling method for pretraining, which provides environmental prior information for the reinforcement learning stage. Subsequently, reinforcement learning is performed within the dynamically generated emotional latent space obtained during pretraining, where the latent representations are further optimized according to the subjects’ real emotional feedback. The integration of unsupervised pretraining and reinforcement learning further improves the emotion prediction performance of the proposed model.

Fig.\ref{fig:Keyframe detection} presents the emotion prediction results of the trained model on the SEED test set. For EEG signals under different emotional categories, the model outputs continuous emotional values at each time step. Fig.\ref{fig:Keyframe detection} (a) illustrates the prediction results for negative sadness emotions. The predicted continuous emotional curve exhibits an overall downward trend accompanied by multiple fluctuations, indicating that the proposed model is capable of identifying temporal turning points of emotional transitions to a certain extent. Fig.\ref{fig:Keyframe detection} (b) presents the prediction results for neutral emotions. The neutral emotional state remains generally stable, with only slight local fluctuations. Nevertheless, the model can still capture subtle variations in weak EEG emotional features and accurately predict minor emotional changes. Fig.\ref{fig:Keyframe detection} (c) shows the prediction results for positive and pleasant emotions. The predicted emotional curve demonstrates an overall upward trend, and the average emotional value is higher than that of neutral emotions. Compared with neutral emotions, positive emotions exhibit larger dynamic fluctuations. Based on variations in EEG signal characteristics, the model accurately predicts emotional transition points and fluctuation ranges, effectively fitting the continuous variations of the ground-truth labels.

Based on the continuous emotion prediction results, this study further extracts emotional key moments. As shown in Fig.\ref{fig:Keyframe detection}, emotional key frames are identified according to the local extrema of the predicted emotional values. Corresponding EEG topographic maps are visualized and compared with those of non-key frames.

Specifically, Fig.\ref{fig:Keyframe detection} (a) presents the prediction curves and EEG topographic maps of key frames for negative emotions. The extracted key frames exhibit significant activations in brain regions associated with sadness-related emotions, whereas the non-key frames show no obvious activation patterns. Fig. \ref{fig:Keyframe detection} (b) illustrates the prediction results and EEG topographic maps for neutral emotions. Although the overall differences in whole-brain activation under neutral emotions are relatively small, the activation intensity of the key frames remains significantly higher than that of the non-key frames. Fig. \ref{fig:Keyframe detection} (c) presents the corresponding results for positive emotions. Compared with neutral emotions, positive emotions exhibit more distinct differences in brain activation patterns. EEG topographic maps at emotional key moments show strong activations in relevant physiological brain regions, while the activation intensity during non-key moments is relatively weak.

In summary, the proposed method not only achieves high-accuracy continuous emotion prediction, but also efficiently extracts emotional key moments from EEG signals based on the continuous emotion prediction results.
}
}

\subsection{{Hyperparameter Sensitivity Analysis}}
{
To analyze the hyperparameter sensitivity of the proposed model, we conducted cross-subject experimental analyses on the SEED and SEED-IV datasets, as illustrated in Figure \ref{fig:Sensitivity analysis of hyperparameters}. Specifically, we investigated three groups of hyperparameters: (1) the number of codebooks, (2) the masking ratio for dynamic emotional latent space modeling, and (3) the maximum sequence length of the Transformer.

(1) Number of Codebooks.
As shown in Fig. \ref{fig:Sensitivity analysis of hyperparameters}(a), the hyperparameter experiments regarding the number of codebooks on the SEED and SEED-IV datasets were conducted with values of 32, 64, 128, and 256. The results indicate that, under the same experimental settings, the best performance was achieved when the number of codebooks was set to 64. These findings suggest that the number of codebooks should be adjusted according to the scale of the dataset, while the optimal value should ultimately be determined empirically.

(2) Masking Ratio for Dynamic Emotional Latent Space Modeling.
As shown in Fig. \ref{fig:Sensitivity analysis of hyperparameters}(b), the hyperparameter experiments on the masking ratio for long-term temporal modeling were conducted on the SEED and SEED-IV datasets with masking ratios of 0.10, 0.25, 0.50, and 0.75. The experimental results demonstrate that, under identical settings, the best performance was obtained when the masking ratio was set to 0.50. Positioned between the 15

(3) Maximum Sequence Length of the Transformer.
As shown in Fig. \ref{fig:Sensitivity analysis of hyperparameters}(c) and Fig. \ref{fig:Sensitivity analysis of hyperparameters}(d), the hyperparameter experiments on the maximum sequence length were conducted on the SEED and SEED-IV datasets by evaluating the MSE and PCC correlations. The maximum sequence lengths were set to 8, 16, 32, 64, 128, and 256, respectively. The experimental results indicate that, under the same parameter settings, the SEED dataset achieved the best performance when the maximum window size was set to 128, whereas the SEED-IV dataset achieved the best performance when the maximum window size was set to 64. Overall, the choice of window size is closely related to the maximum length of an episode or the entire EEG sequence. To a certain extent, larger window sizes generally lead to better performance.
}

\section{Conclusion}
\label{subsec:Conclusion}{
{This paper proposes a novel EEG emotion temporal prediction model, termed EEGDancer, to address the critical challenge of capturing fine-grained dynamic emotional changes from EEG signals. Unlike conventional deep learning approaches, EEGDancer is the first framework to formulate EEG emotion temporal prediction as a Markov Decision Process (MDP), where the prediction process is optimized through a reinforcement learning reward mechanism. Different from traditional supervised learning methods that only focus on prediction accuracy at the current time step, the reward mechanism enables the model not only to consider the current prediction outcome, but also to account for how the current decision influences future dynamic prediction results. Consequently, the optimization process is transformed from a sample-to-sample local optimization into a global trajectory optimization, further aligning sequential prediction decisions with the evolving trends of EEG feature dynamics over an entire temporal sequence.

Similar to how a dancer coordinates with the rhythm and tempo of music to perform graceful and expressive movements, the proposed model learns the intrinsic rhythmic and temporal patterns of EEG signals, while enabling the prediction results to more accurately track the underlying dynamic evolution of human emotions.

Extensive experimental results conducted on three benchmark datasets demonstrate that EEGDancer achieves SOTA performance in cross-subject EEG emotion temporal prediction tasks. Comprehensive ablation studies and hyperparameter sensitivity analyses further verify the superiority and robustness of the EEGDancer framework. Feature visualization results reveal that the model compresses long-term EEG emotional dynamics into a manifold distribution, making the directional evolution of EEG emotional features temporally continuous and predictable, and further capture the intrinsic continuous evolutionary patterns of EEG emotions through a reinforcement learning-based reward mechanism
, thereby enhancing the temporal prediction capability of the model. These findings demonstrate that EEGDancer has strong potential for practical deployment in aBCI applications.
}
}

\section{Acknowledgments}
This work was supported by the National Natural Science Foundation of China (62276169), Medical-Engineering Interdisciplinary Research Foundation of Shenzhen University (2024YG008), Shenzhen University-Lingnan University Joint Research Programme, Shenzhen-Hong Kong Institute of Brain Science-Shenzhen Fundamental Research Institutions (2023SHIBS0003), the STI 2030-Major Projects (2021ZD0200500), the Open Research Fund of the State Key Laboratory of Brain-Machine Intelligence, Zhejiang University (Grant No. BMI2400008), and the Shenzhen Science and Technology Program (No. JCYJ20241202124222027 and JCYJ20241202124209011).


\bibliographystyle{./elsarticle-num.bst}
\bibliography{./main.bib}

@article{li2022eeg,
  title={EEG based emotion recognition: A tutorial and review},
  author={Li, Xiang and Zhang, Yazhou and Tiwari, Prayag and Song, Dawei and Hu, Bin and Yang, Meihong and Zhao, Zhigang and Kumar, Neeraj and Marttinen, Pekka},
  journal={ACM Computing Surveys},
  volume={55},
  number={4},
  pages={1--57},
  year={2022},
  publisher={ACM New York, NY}
}

@article{alarcao2017emotions,
  title={Emotions recognition using EEG signals: A survey},
  author={Alarcao, Soraia M and Fonseca, Manuel J},
  journal={IEEE transactions on affective computing},
  volume={10},
  number={3},
  pages={374--393},
  year={2017},
  publisher={IEEE}
}

@article{houssein2022human,
  title={Human emotion recognition from EEG-based brain--computer interface using machine learning: a comprehensive review},
  author={Houssein, Essam H and Hammad, Asmaa and Ali, Abdelmgeid A},
  journal={Neural Computing and Applications},
  volume={34},
  number={15},
  pages={12527--12557},
  year={2022},
  publisher={Springer}
}

@article{peng2022ogssl,
  title={OGSSL: A semi-supervised classification model coupled with optimal graph learning for EEG emotion recognition},
  author={Peng, Yong and Jin, Fengzhe and Kong, Wanzeng and Nie, Feiping and Lu, Bao-Liang and Cichocki, Andrzej},
  journal={IEEE Transactions on Neural Systems and Rehabilitation Engineering},
  volume={30},
  pages={1288--1297},
  year={2022},
  publisher={IEEE}
}

@article{pancholi2022source,
  title={Source aware deep learning framework for hand kinematic reconstruction using EEG signal},
  author={Pancholi, Sidharth and Giri, Amita and Jain, Anant and Kumar, Lalan and Roy, Sitikantha},
  journal={IEEE Transactions on Cybernetics},
  volume={53},
  number={7},
  pages={4094--4106},
  year={2022},
  publisher={IEEE}
}

@misc{zoph2017neuralarchitecturesearchreinforcement,
      title={Neural Architecture Search with Reinforcement Learning}, 
      author={Barret Zoph and Quoc V. Le},
      year={2017},
      eprint={1611.01578},
      archivePrefix={arXiv},
}

@misc{he2016deepreinforcementlearningnatural,
      title={Deep Reinforcement Learning with a Natural Language Action Space}, 
      author={Ji He and Jianshu Chen and Xiaodong He and Jianfeng Gao and Lihong Li and Li Deng and Mari Ostendorf},
      year={2016},
      eprint={1511.04636},
      archivePrefix={arXiv},
}

@inproceedings{yarats2021image,
  title={Image augmentation is all you need: Regularizing deep reinforcement learning from pixels},
  author={Yarats, Denis and Kostrikov, Ilya and Fergus, Rob},
  booktitle={International conference on learning representations},
  year={2021}
}

@article{yang2021cnn,
  title={A CNN identified by reinforcement learning-based optimization framework for EEG-based state evaluation},
  author={Yang, Yuxuan and Gao, Zhongke and Li, Yanli and Wang, He},
  journal={Journal of Neural Engineering},
  volume={18},
  number={4},
  pages={046059},
  year={2021},
  publisher={IOP Publishing}
}

@article{zhang2023unsupervised,
  title={Unsupervised time-aware sampling network with deep reinforcement learning for eeg-based emotion recognition},
  author={Zhang, Yongtao and Pan, Yue and Zhang, Yulin and Zhang, Min and Li, Linling and Zhang, Li and Huang, Gan and Su, Lei and Liang, Zhen and Zhang, Zhiguo},
  journal={IEEE Transactions on Affective Computing},
  year={2023},
  publisher={IEEE}
}

@article{aung2025real,
  title={A Real-Time Framework for EEG Signal Decoding With Graph Neural Networks and Reinforcement Learning},
  author={Aung, Htoo Wai and Li, Jiao Jiao and An, Yang and Su, Steven Weidong},
  journal={IEEE Transactions on Neural Networks and Learning Systems},
  year={2025},
  publisher={IEEE}
}

@article{liu2023automatic,
  title={Automatic focal EEG identification based on deep reinforcement learning},
  author={Liu, Xinyu and Ding, Xin and Liu, Jianping and Nie, Weiwei and Yuan, Qi},
  journal={Biomedical Signal Processing and Control},
  volume={83},
  pages={104693},
  year={2023},
  publisher={Elsevier}
}

@article{xu2021accelerating,
  title={Accelerating reinforcement learning using eeg-based implicit human feedback},
  author={Xu, Duo and Agarwal, Mohit and Gupta, Ekansh and Fekri, Faramarz and Sivakumar, Raghupathy},
  journal={Neurocomputing},
  volume={460},
  pages={139--153},
  year={2021},
  publisher={Elsevier}
}

@inproceedings{luo2018deep,
  title={Deep reinforcement learning from error-related potentials via an EEG-based brain-computer interface},
  author={Luo, Tian-jian and Fan, Ya-chao and Lv, Ji-tu and Zhou, Chang-le},
  booktitle={2018 IEEE international conference on bioinformatics and biomedicine (BIBM)},
  pages={697--701},
  year={2018},
  organization={IEEE}
}

@article{li2023brain,
  title={Brain emotion perception inspired EEG emotion recognition with deep reinforcement learning},
  author={Li, Dongdong and Xie, Li and Wang, Zhe and Yang, Hai},
  journal={IEEE Transactions on Neural Networks and Learning Systems},
  year={2023},
  publisher={IEEE}
}

@article{zheng2015investigating,
  title={Investigating critical frequency bands and channels for EEG-based emotion recognition with deep neural networks},
  author={Zheng, Wei-Long and Lu, Bao-Liang},
  journal={IEEE Transactions on autonomous mental development},
  volume={7},
  number={3},
  pages={162--175},
  year={2015},
  publisher={IEEE}
}

@article{zheng2018emotionmeter,
  title={Emotionmeter: A multimodal framework for recognizing human emotions},
  author={Zheng, Wei-Long and Liu, Wei and Lu, Yifei and Lu, Bao-Liang and Cichocki, Andrzej},
  journal={IEEE transactions on cybernetics},
  volume={49},
  number={3},
  pages={1110--1122},
  year={2018},
  publisher={IEEE}
}

@article{hu2023eeg,
  title={EEG microstate correlates of emotion dynamics and stimulation content during video watching},
  author={Hu, Wanrou and Zhang, Zhiguo and Zhao, Huilin and Zhang, Li and Li, Linling and Huang, Gan and Liang, Zhen},
  journal={Cerebral Cortex},
  volume={33},
  number={3},
  pages={523--542},
  year={2023},
  publisher={Oxford University Press}
}

@article{hu2022microstate,
  title={Microstate detection in naturalistic electroencephalography data: A systematic comparison of topographical clustering strategies on an emotional database},
  author={Hu, Wanrou and Zhang, Zhiguo and Zhang, Li and Huang, Gan and Li, Linling and Liang, Zhen},
  journal={Frontiers in Neuroscience},
  volume={16},
  pages={812624},
  year={2022},
  publisher={Frontiers Media SA}
}

@article{drucker1996support,
  title={Support vector regression machines},
  author={Drucker, Harris and Burges, Christopher J and Kaufman, Linda and Smola, Alex and Vapnik, Vladimir},
  journal={Advances in neural information processing systems},
  volume={9},
  year={1996}
}

@article{cover1967nearest,
  title={Nearest neighbor pattern classification},
  author={Cover, Thomas and Hart, Peter},
  journal={IEEE transactions on information theory},
  volume={13},
  number={1},
  pages={21--27},
  year={1967},
  publisher={IEEE}
}

@article{robinson1965machine,
  title={A machine-oriented logic based on the resolution principle},
  author={Robinson, John Alan},
  journal={Journal of the ACM (JACM)},
  volume={12},
  number={1},
  pages={23--41},
  year={1965},
  publisher={ACM New York, NY, USA}
}

@article{breiman2001random,
  title={Random forests},
  author={Breiman, Leo},
  journal={Machine learning},
  volume={45},
  number={1},
  pages={5--32},
  year={2001},
  publisher={Springer}
}

@article{lindley1972bayes,
  title={Bayes estimates for the linear model},
  author={Lindley, Dennis V and Smith, Adrian FM},
  journal={Journal of the Royal Statistical Society Series B: Statistical Methodology},
  volume={34},
  number={1},
  pages={1--18},
  year={1972},
  publisher={Oxford University Press}
}

@article{friedman2001greedy,
  title={Greedy function approximation: a gradient boosting machine},
  author={Friedman, Jerome H},
  journal={Annals of statistics},
  pages={1189--1232},
  year={2001},
  publisher={JSTOR}
}

@article{tibshirani1996regression,
  title={Regression shrinkage and selection via the lasso},
  author={Tibshirani, Robert},
  journal={Journal of the Royal Statistical Society Series B: Statistical Methodology},
  volume={58},
  number={1},
  pages={267--288},
  year={1996},
  publisher={Oxford University Press}
}

@article{hoerl1970ridge,
  title={Ridge regression: Biased estimation for nonorthogonal problems},
  author={Hoerl, Arthur E and Kennard, Robert W},
  journal={Technometrics},
  volume={12},
  number={1},
  pages={55--67},
  year={1970},
  publisher={Taylor \& Francis}
}

@article{rumelhart1986learning,
  title={Learning representations by back-propagating errors},
  author={Rumelhart, David E and Hinton, Geoffrey E and Williams, Ronald J},
  journal={nature},
  volume={323},
  number={6088},
  pages={533--536},
  year={1986},
  publisher={Nature Publishing Group UK London}
}

@incollection{huber1992robust,
  title={Robust estimation of a location parameter},
  author={Huber, Peter J},
  booktitle={Breakthroughs in statistics: Methodology and distribution},
  pages={492--518},
  year={1992},
  publisher={Springer}
}

@article{lawhern2018eegnet,
  title={EEGNet: a compact convolutional neural network for EEG-based brain--computer interfaces},
  author={Lawhern, Vernon J and Solon, Amelia J and Waytowich, Nicholas R and Gordon, Stephen M and Hung, Chou P and Lance, Brent J},
  journal={Journal of neural engineering},
  volume={15},
  number={5},
  pages={056013},
  year={2018},
  publisher={iOP Publishing}
}

@inproceedings{li2018novel,
  title={A novel neural network model based on cerebral hemispheric asymmetry for EEG emotion recognition.},
  author={Li, Yang and Zheng, Wenming and Cui, Zhen and Zhang, Tong and Zong, Yuan},
  booktitle={IJCAI},
  pages={1561--1567},
  year={2018}
}

@article{chen2021ms,
  title={MS-MDA: Multisource marginal distribution adaptation for cross-subject and cross-session EEG emotion recognition},
  author={Chen, Hao and Jin, Ming and Li, Zhunan and Fan, Cunhang and Li, Jinpeng and He, Huiguang},
  journal={Frontiers in Neuroscience},
  volume={15},
  pages={778488},
  year={2021},
  publisher={Frontiers Media SA}
}

@inproceedings{sun2016return,
  title={Return of frustratingly easy domain adaptation},
  author={Sun, Baochen and Feng, Jiashi and Saenko, Kate},
  booktitle={Proceedings of the AAAI conference on artificial intelligence},
  volume={30},
  number={1},
  year={2016}
}

@article{zhong2020eeg,
  title={EEG-based emotion recognition using regularized graph neural networks},
  author={Zhong, Peixiang and Wang, Di and Miao, Chunyan},
  journal={IEEE Transactions on Affective Computing},
  volume={13},
  number={3},
  pages={1290--1301},
  year={2020},
  publisher={IEEE}
}

@inproceedings{li2018cross,
  title={Cross-subject emotion recognition using deep adaptation networks},
  author={Li, He and Jin, Yi-Ming and Zheng, Wei-Long and Lu, Bao-Liang},
  booktitle={International conference on neural information processing},
  pages={403--413},
  year={2018},
  organization={Springer}
}

@article{ganin2016domain,
  title={Domain-adversarial training of neural networks},
  author={Ganin, Yaroslav and Ustinova, Evgeniya and Ajakan, Hana and Germain, Pascal and Larochelle, Hugo and Laviolette, Fran{\c{c}}ois and March, Mario and Lempitsky, Victor},
  journal={Journal of machine learning research},
  volume={17},
  number={59},
  pages={1--35},
  year={2016}
}

@article{zhou2025emotion,
  title={Emotion agent: Unsupervised deep reinforcement learning with distribution-prototype reward for continuous emotional EEG analysis},
  author={Zhou, Zhihao and Zhang, Li and Liu, Qile and Huang, Gan and Yu, Zhuliang and Liang, Zhen},
  journal={Neurocomputing},
  pages={130951},
  year={2025},
  publisher={Elsevier}
}

@article{guo2025deepseek,
  title={DeepSeek-R1 incentivizes reasoning in LLMs through reinforcement learning},
  author={Guo, Daya and Yang, Dejian and Zhang, Haowei and Song, Junxiao and Wang, Peiyi and Zhu, Qihao and Xu, Runxin and Zhang, Ruoyu and Ma, Shirong and Bi, Xiao and others},
  journal={Nature},
  volume={645},
  number={8081},
  pages={633--638},
  year={2025},
  publisher={Nature Publishing Group UK London}
}

@article{elad2006image,
  title={Image denoising via sparse and redundant representations over learned dictionaries},
  author={Elad, Michael and Aharon, Michal},
  journal={IEEE Transactions on Image processing},
  volume={15},
  number={12},
  pages={3736--3745},
  year={2006},
  publisher={IEEE}
}

@article{zhang2025mind,
  title={MIND-EEG: Multi-granularity Integration Network with Discrete Codebook for EEG-based Emotion Recognition},
  author={Zhang, Yuzhe and Xie, Chengxi and Liu, Huan and Shi, Yuhan and Liu, Guanjian and Zhang, Dalin},
  journal={IEEE Transactions on Affective Computing},
  year={2025},
  publisher={IEEE}
}

@inproceedings{ren2025videoworld,
  title={Videoworld: Exploring knowledge learning from unlabeled videos},
  author={Ren, Zhongwei and Wei, Yunchao and Guo, Xun and Zhao, Yao and Kang, Bingyi and Feng, Jiashi and Jin, Xiaojie},
  booktitle={Proceedings of the Computer Vision and Pattern Recognition Conference},
  pages={29029--29039},
  year={2025}
}

@article{yang2010image,
  title={Image super-resolution via sparse representation},
  author={Yang, Jianchao and Wright, John and Huang, Thomas S and Ma, Yi},
  journal={IEEE transactions on image processing},
  volume={19},
  number={11},
  pages={2861--2873},
  year={2010},
  publisher={IEEE}
}

@article{gray1984vector,
  title={Vector quantization},
  author={Gray, Robert},
  journal={IEEE Assp Magazine},
  volume={1},
  number={2},
  pages={4--29},
  year={1984},
  publisher={IEEE}
}

@article{makhoul1985vector,
  title={Vector quantization in speech coding},
  author={Makhoul, John and Roucos, Salim and Gish, Herbert},
  journal={Proceedings of the IEEE},
  volume={73},
  number={11},
  pages={1551--1588},
  year={1985},
  publisher={IEEE}
}

@article{mao2021discrete,
  title={Discrete representations strengthen vision transformer robustness},
  author={Mao, Chengzhi and Jiang, Lu and Dehghani, Mostafa and Vondrick, Carl and Sukthankar, Rahul and Essa, Irfan},
  journal={arXiv preprint arXiv:2111.10493},
  year={2021}
}

@inproceedings{aczel2026neural,
  title={Neural Audio Compression without Residual Vector Quantization},
  author={Aczel, Till and Lanzend{\"o}rfer, Luca A and Gao, Fei and Wattenhofer, Roger},
  booktitle={AAAI 2026 Workshop on Machine Learning for Wireless Communication and Networks (ML4Wireless)}
}

@inproceedings{zhao2026spherical,
  title={Spherical leech quantization for visual tokenization and generation},
  author={Zhao, Yue and Jiang, Hanwen and Xu, Zhenlin and Yang, Chutong and Adeli, Ehsan and Kr{\"a}henb{\"u}hl, Philipp},
  booktitle={Proceedings of the IEEE/CVF Conference on Computer Vision and Pattern Recognition},
  pages={12913--12923},
  year={2026}
}

@inproceedings{haarnoja2018soft,
  title={Soft actor-critic: Off-policy maximum entropy deep reinforcement learning with a stochastic actor},
  author={Haarnoja, Tuomas and Zhou, Aurick and Abbeel, Pieter and Levine, Sergey},
  booktitle={International conference on machine learning},
  pages={1861--1870},
  year={2018},
  organization={Pmlr}
}

@inproceedings{fujimoto2018addressing,
  title={Addressing function approximation error in actor-critic methods},
  author={Fujimoto, Scott and Hoof, Herke and Meger, David},
  booktitle={International conference on machine learning},
  pages={1587--1596},
  year={2018},
  organization={PMLR}
}

@article{schulman2017proximal,
  title={Proximal policy optimization algorithms},
  author={Schulman, John and Wolski, Filip and Dhariwal, Prafulla and Radford, Alec and Klimov, Oleg},
  journal={arXiv preprint arXiv:1707.06347},
  year={2017}
}

@inproceedings{silver2014deterministic,
  title={Deterministic policy gradient algorithms},
  author={Silver, David and Lever, Guy and Heess, Nicolas and Degris, Thomas and Wierstra, Daan and Riedmiller, Martin},
  booktitle={International conference on machine learning},
  pages={387--395},
  year={2014},
  organization={Pmlr}
}

@article{shu2018review,
  title={A review of emotion recognition using physiological signals},
  author={Shu, Lin and Xie, Jinyan and Yang, Mingyue and Li, Ziyi and Li, Zhenqi and Liao, Dan and Xu, Xiangmin and Yang, Xinyi},
  journal={Sensors},
  volume={18},
  number={7},
  pages={2074},
  year={2018},
  publisher={MDPI}
}

@article{ji2022spatial,
  title={Spatial-temporal network for fine-grained-level emotion EEG recognition},
  author={Ji, Youshuo and Li, Fu and Fu, Boxun and Li, Yang and Zhou, Yijin and Niu, Yi and Zhang, Lijian and Chen, Yuanfang and Shi, Guangming},
  journal={Journal of Neural Engineering},
  volume={19},
  number={3},
  pages={036017},
  year={2022},
  publisher={IOP Publishing}
}

@article{assemlali2025deep,
  title={Deep learning-driven CNN model for detection and classification of dynamic obstacles},
  author={Assemlali, Hamza and Bouhsissin, Soukaina and Sael, Nawal},
  journal={Green Energy and Intelligent Transportation},
  pages={100334},
  year={2025},
  publisher={Elsevier}
}

@article{atlas2025modernized,
  title={A modernized approach to sentiment analysis of product reviews using BiGRU and RNN based LSTM deep learning models},
  author={Atlas, L Godlin and Arockiam, Daniel and Muthusamy, Arvindhan and Balusamy, Balamurugan and Selvarajan, Shitharth and Al-Shehari, Taher and Alsadhan, Nasser A},
  journal={Scientific Reports},
  volume={15},
  number={1},
  pages={16642},
  year={2025},
  publisher={Nature Publishing Group UK London}
}

@article{yu2026ia,
  title={IA 2 GNN: Imbalance-Aware Adaptive Graph Construction for Multi-Modal Image Fusion},
  author={Yu, Dong and Tang, Yepeng and Zhang, Chunjie and Wang, Wei and Yang, Guodong and Zheng, Xiaolong and Zhao, Yao},
  journal={IEEE Transactions on Multimedia},
  year={2026},
  publisher={IEEE}
}

@article{wang2025cross,
  title={Cross-dataset EEG emotion recognition based on pre-trained Vision Transformer considering emotional sensitivity diversity},
  author={Wang, Fang and Tian, Yu-Chu and Zhou, Xiaobo},
  journal={Expert Systems with Applications},
  volume={279},
  pages={127348},
  year={2025},
  publisher={Elsevier}
}

@article{luo2026multi,
  title={Multi-scale cross-domain and class-wise kernel discriminative alignment for EEG-based emotion recognition},
  author={Luo, Gang and Zheng, Chengcheng and Zhu, Lixian and Fan, Tianqi and Tian, Fuze and Wang, Dixin and Qian, Kun and Liu, Jingxin and Sun, Shuting and Hu, Bin},
  journal={Pattern Recognition},
  pages={113626},
  year={2026},
  publisher={Elsevier}
}

@article{liu2026cria,
  title={Cria: A cross-view interaction and instance-adapted pre-training framework for generalizable eeg representations},
  author={Liu, Puchun and Chen, CL Philip and He, Yubin and Zhang, Tong},
  journal={Pattern Recognition},
  pages={113272},
  year={2026},
  publisher={Elsevier}
}

@inproceedings{zhou2024emotvr,
  title={EMOTVR: a hybrid model to estimate continuous-time and continuous-level emotion from electroencephalography},
  author={Zhou, Xinxu and Liang, Zhen and Ye, Weishan and Xue, Junqi and Liu, Honghai and Zhang, Min and Zhang, Zhiguo},
  booktitle={ICASSP 2024-2024 IEEE International Conference on Acoustics, Speech and Signal Processing (ICASSP)},
  pages={2021--2025},
  year={2024},
  organization={IEEE}
}

@inproceedings{ma2025codebrain,
  title={Codebrain: Bridging decoupled tokenizer and multi-scale architecture for eeg foundation model},
  author={Ma, Jingying and Wu, Feng and Lin, Qika and Xing, Yucheng and Liu, Chenyu and Jia, Ziyu and Feng, Mengling},
  booktitle={The Fourteenth International Conference on Learning Representations},
  year={2025}
}

@article{bao2021beit,
  title={Beit: Bert pre-training of image transformers},
  author={Bao, Hangbo and Dong, Li and Piao, Songhao and Wei, Furu},
  journal={arXiv preprint arXiv:2106.08254},
  year={2021}
}

@article{ju2024eeg,
  title={EEG-based emotion recognition using a temporal-difference minimizing neural network},
  author={Ju, Xiangyu and Li, Ming and Tian, Wenli and Hu, Dewen},
  journal={Cognitive Neurodynamics},
  volume={18},
  number={2},
  pages={405--416},
  year={2024},
  publisher={Springer}
}

@inproceedings{liu2025eeg,
  title={Eeg-scmm: Soft contrastive masked modeling for cross-corpus eeg-based emotion recognition},
  author={Liu, Qile and Ye, Weishan and Zhang, Lingli and Liang, Zhen},
  booktitle={Proceedings of the 33rd ACM International Conference on Multimedia},
  pages={5834--5842},
  year={2025}
}

\end{document}